\documentclass[sigconf]{acmart}
\usepackage{blkarray}
\usepackage{amsmath,amsfonts}
\usepackage[linesnumbered,ruled,vlined]{algorithm2e}
\usepackage{caption,subfig}
\usepackage{gensymb}

\usepackage[font=footnotesize]{caption}
\usepackage{graphicx}
\usepackage{multirow}
\graphicspath{{./images/}}
\usepackage{textcomp}

\def\BibTeX{{\rm B\kern-.05em{\sc i\kern-.025em b}\kern-.08em
    T\kern-.1667em\lower.7ex\hbox{E}\kern-.125emX}}
\AtBeginDocument{%
  \providecommand\BibTeX{{%
    \normalfont B\kern-0.5em{\scshape i\kern-0.25em b}\kern-0.8em\TeX}}}

\setcopyright{acmcopyright}
\copyrightyear{2018}
\acmYear{2018}
\acmDOI{10.1145/1122445.1122456}

\acmConference[Woodstock '18]{Woodstock '18: ACM Symposium on Neural
  Gaze Detection}{June 03--05, 2018}{Woodstock, NY}
\acmBooktitle{Woodstock '18: ACM Symposium on Neural Gaze Detection,
  June 03--05, 2018, Woodstock, NY}
\acmPrice{15.00}
\acmISBN{978-1-4503-XXXX-X/18/06}

\begin{document}

\title{Digital Contact Tracing for Covid-19}

 \author{Chandresh Kumar Maurya, Seemandhar Jain,  Vishal Thakre}
 \email{{Chandresh,cse170001046,cse180001062}@iiti.ac.in}
 \orcid{1234-5678-9012}
 %\author{Seemandhar Jain}
 %\email{cse170001046@iiti.ac.in}
 \affiliation{%
   \institution{Indian Institute of Technology Indore}
   %\streetaddress{P.O. Box 1212}
   \city{Indore}
   \state{Madhya Pradesh}
   \country{India}
   \postcode{452020}
 }

\renewcommand{\shortauthors}{Trovato and Tobin, et al.}

\begin{abstract}
 Recently, the COVID-19 pandemic created a worldwide emergency as it is estimated that such a large number of infections are due to human-to-human transmission of the COVID-19. As a necessity, there is a need to track users who came in contact with users having travel history, asymptomatic and not yet symptomatic, but they can be in the future. To solve this problem, the present work proposes a solution for contact tracing based on assisted GPS and cloud computing technologies. An application is developed to collect each user's assisted GPS coordinates once all the users install this application. This application periodically sends assisted GPS data (coordinates) to the cloud. To determine which devices are within the permissible limit of 5m (tunable parameter),  we perform clustering over assisted GPS coordinates and track the clusters for about $t$ mins (tunable parameter) to allow the measure of spread. We assume that it takes around 3-5 mins to get the virus from an infected object. For clustering, the proposed M-way like tree data structure stores the assisted GPS coordinates in degree, minute, and second (DMS) format. Thus, every user is mapped to a  leaf node of the tree. The crux of the solution lies at the leaf node. We split the "seconds" part of the assisted GPS location into $m$ equal parts (a tunable parameter), which amount to $d$ meter in latitude/longitude. Hence, two users who are within $d$ meter range will map to the same leaf node. Thus, by mapping assisted GPS locations every $t$ mins (usually $t = 2.5$ mins), we can find out how many users came in contact with a particular user for at least $t$ mins. Our work's salient feature is that it runs in linear time $O(n)$ for $n$ users in the static case, i.e., when users are not moving. We also propose a variant of our solution to handle the dynamic case, that is, when users are moving. Besides, the proposed solution offers \emph{potential hotspot detection} and \emph{safe-route recommendation} as an additional feature, and proof-of-concept is presented through experiments on simulated data of 2/4/6/8/10M users. 
\end{abstract}


\maketitle
\section{Introduction}
Contact tracing is the problem of identifying users who have come in contact with a user within a certain distance for a specific amount of time. This problem came into the limelight during the COVID-19 pandemic, which is thought to spread from humans to humans. The main problem is that users are asymptomatic and can pass the virus to other users unknowingly for weeks or months. In such a case, it is challenging to identify users who came in contact with a particular user, given the users dynamic and complex movement patterns. Another prevalent problem during the pandemic situation is that users are interested in knowing the hotspot areas to avoid them while visiting. Therefore, there is a growing need for a recommendation of a safe route for travel. 

Current solutions for contact tracing problems can be categorized into (a) human-based approach, (b) Bluetooth (BT)-based approach, and (c) GPS-based approach. In the first approach, police personnel are deployed to follow the user's movement trail and find out users who were present in close proximity of the particular user for a specific duration. This approach is \emph{costly, cumbersome, and time-consuming} \cite{covidappsurvey}. In the second approach, each user is asked to install an app such as India's Aarogya Setu app or Google-Apple privacy-preserving app. Such apps use BT signals for exchanging data between devices. There are certain limitations of these apps. For example, the limitations of Aarogya Setu are: (a) it does not tell if we were in contact with a user a week or month ago who recently got diagnosed with COVID-19, (b) it also does not tell which areas are hotspots and should be avoided, (c) it is based on self-assessment only, and hence reliability is a concern. 

The solutions for hotspot detection are currently based on the government's rules and regulations. For example, sites like COVID tracker \footnote{https://www.COVIDhotspots.in/?city=Delhi} mark geographical areas as hotspot-based on the directions received from the government. They do not provide real-time hotspot information, such as alerting the user that they are moving through hotspot/containment zones specially in villages where monitoring is difficult. Further, the existing solutions lack the feature of recommending safe routes for travel, such as Google map.

To overcome the limitations of the existing solutions, we propose an assisted GPS location-based solution that receives the assisted GPS coordinates of the user in the backend and without forcing the user to keep on Bluetooth time and again. This solution does not require any user interventions and sends the assisted GPS coordinates to the cloud periodically. Received coordinates are mapped to the proposed {\bf  M-way like tree data structure} in degree, minute, and second (DMS) format. Once the mapping is over, the proposed clustering algorithm executes to track the history of users who spent at least $t$ minutes with other users.  {\bf  Our solution is scalable and can find users who happen to make a contact event in 17 mins among a population of 10M users.}

Major differences between our solution based on assisted GPS and BT-based solution for contact tracing are: (i) our approach can provide location information where the user got {\bf potentially infected} whereas BT-based approaches fail to do so since they do not provide location information. Why this information may be crucial is that let us suppose that the user visits a dairy milk shop every day besides tons of other places. On one day (s)he got COVID +ve. In this case, (s)he might be interested in knowing where (s)he got the virus so that they can inform their family members to be vigilant. In this scenario, the BT-based approach has no clue of the location, (ii) assisted GPS-based solution is centralized where BT-based one is decentralized, (iii) our solution is more robust in the sense that BT-based solution communicates with {\bf many heterogeneous devices}.  In contrast, ours does not, (iv) BT always needs to be turned on, which usually user forget or do not turn on for power saving purposes. As for assisted GPS is concerned, once the app is installed, it will keep using the location sharing in the background and does not ask the user to turn it on time and again. This option is not available in BT-based solutions currently, (v) the scalability of BT-based apps is also a concern since these apps can communicate to only eight other BT devices at a time. Further, note that we are using \emph{assisted} GPS which is more reliable than GPS, where location information is calculated using satellite signals, mobile towers, and wifi beacons. Therefore, our solution will work in indoor settings as well. From now on, whenever GPS is mentioned will mean assisted GPS. To emphasize to our readers that our solution offers an alternative to BT-based solutions and does not replace them.

In short, we make the following contributions:
\begin{itemize}
\item Propose a solution for contact tracing that runs in the time linear to the numbers of users i.e. $O(n)$. the space complexity of the solution is also linear in the number of user identifiers, i.e., $O(n)$.
\item Our solution can handle static as well as dynamic case. That means if users stay at the same location (static case) or the users are moving together (dynamic case).
\item The proposed solution relies only on location sharing information, which is the property of most android applications such as Google news, maps, etc.
\item Our solution can additionally provide hotspot information (areas of large gatherings) and recommend a safe route for travel.
\end{itemize}

\section{Related Works}
Contact tracing and regular monitoring of infectious diseases (COVID-19) are essential for public health. Utilizing emerging technologies such as remote sensing, internet-based surveillance, telecommunications, infectious disease modeling, Global Positioning System (GPS), IoT devices, and mobile phones, such contagious diseases can be monitored, prevented (by generating alarms after real-time predictions), and controlled \cite{christaki2015new}. 
New approaches to deal with such epidemics are recently categorized with the newly coined term such as digital epidemiology \cite{salathe2012digital}. Some works have appraised the characterization of the human mobility patterns by using Call Detail Records (CDRs) for understanding and modeling epidemics \cite{cdr}. The authors discovered the opportunity to utilize individual mobility proxies to describe commuting flows and predict influenza-like illness diffusion. Though depending on the human mobility data source, their predictive accuracy about epidemic invasion timing and propagation of patterns differed. 
% One more approach for detecting, tracking, and monitoring contacts utilizes wireless sensor network (WSN) technologies, i.e., Bluetooth. One of the first experiments using WSN is implemented by Salathé et al. \cite{salathe2012digital}. The author acquires high-resolution data of interpersonal mobile contacts on one classic day at the American high school, making it conceivable to reconstruct the pertinent social network from infectious disease transmission. This work also includes an SEIR (Susceptible, Exposed, Infectious, Recovered) model to estimate the disease diffusion and study the impact of vaccination measures.

Mastrandrea et al. \cite{mastrandrea2015contact} utilizes wearable sensors to collect contacts among students and compare the results with the personal diaries' contacts. The authors also compare how such epidemic diseases spread using two different contact networks (from diaries and sensors), which displays a notable alteration in dynamics. Contact tracing is one of the very useful measures which primarily focuses on potential next-generation cases. It has been proved a highly successful strategy if the numbers of infectious epidemic cases are low, or at early stages of the outbreak, mainly, the disease is asymptomatic (still infectious), because it offers a way by which such individuals are easily recognized.  Broadly, there can be two approaches for modeling contact tracing \cite{kwok2019epidemic}: (1) Population-based modeling is the top-down approach used to depict the disease dynamics on the system level, which is typically used to analyze research related matters from a macroscopic perspective. (2) agent-based modeling, which is a bottom-up technique and deals with every individual agent, each with their infection states and own movements. It is usually used to evaluate adaptive and heterogeneous behaviors.  The latter approach is more realistic. However, it can be computationally expensive.
Few research works \cite{ascione2020construction, muller2000contact}  have introduced the stochastic model used to reduce a deterministic approach for attaining the fundamental dynamics of the epidemics and other associated measures. Mostly, previous models deal only with generic networks. To measure the precision of the trace contact models, there is a need to account for the network of contacts. For example, Huerta and Tsimring \cite{huerta2002contact} present a stochastic model to estimate the effect of random screening and contact tracing as part of the epidemic control strategy in complex networks. This work indicates that a significant outbreak can significantly be reduced via tracing contacts at a low additional cost. A similar approach is also used by Farrahi et al. \cite{farrahi2014epidemic}. 

Ferretti et al. \cite{ferretti2020quantifying} state that isolation and contact tracing are being practiced but not preventing the COVID-19, which is why the same high number of asymptomatic infected individuals remain undetected, and cases continue to spread. Therefore, the authors propose mobile apps to trace previous mobile contacts and show mathematically that epidemics can remain even when not all population use the application. Hellewell et al. \cite{hellewell2020feasibility}  also propose similar inference through the simulated model. In most scenarios, highly effective case isolation and contact tracing are sufficient to control a new outbreak of COVID-19 within three months; however, only 79\% of the contacts are traced. Nevertheless, such conditions create smartphone-based tracing, which is far from a realistic solution.
One of the first efforts through mobile phones to trace the contacts was the FluPhone application developed by Cambridge University \cite{yoneki2011fluphone} which uses Bluetooth-based wireless signals to estimate the physical contact. It also asks the users to report on flu-like symptoms to appraise the risk of infections. Next, the Singapore Government also developed a mobile app called Trace-Together for COVID-19 \footnote{https://www.tracetogether.gov.sg/}. This also uses Bluetooth technology, and it had already been utilized to control the disease spread. Few similar works also focused on privacy as the Pan-European Privacy-Preserving Proximity Tracing (PEPP-PT) \cite{team2020pan}. Finally, Google and Apple have teamed up to design, develop and integrate a similar approach into the Android operating
Systems and iOS. The proposed solution is more efficient and ubiquitous to users. Aarogya Setu app \footnote{https://www.mygov.in/aarogya-Setu-app/} from the Indian government works on location sharing and Bluetooth. Secondly, it tells us that we are in contact with a COVID +ve person based on Bluetooth proximity. There are certain limitations of Aarogya Setu. For example, it does not tell if we were in contact with a person a week or month ago who recently got diagnosed with COVID-19. It also does not reveal which areas are hotspots and should be avoided. It is based on self-assessment only, and hence reliability is a concern. Recently, Mahapatra et al. \cite{mahapatra2020dynamic} present a digital contact tracing solution based on a dynamic graph streaming algorithm. Concretely, they use an index-based adjacency list to store graphs whose nodes are users and edges are close contacts. However, they do not mention if they use Bluetooth or GPS to find direct and indirect close contacts. The proposed solution runs in $O(q^L|I|)$ where $q, L, |I|$ denote the number of contacts, level of tracing, and the number of infected users, respectively.
All Bluetooth (even GPS) based apps for contact tracing suffer from a severe drawback: they may be subjected to high "false alarm". For example, when two users are standing opposite side of a wall, Bluetooth-based contact tracing apps will signal that they are within the infection-proximity \cite{VAUGHAN20209}. There have been reports of delayed notification (1+ days) to individuals regarding contacts with COVID +ve patients. Such delays can create depression in users later on. Further, many of these apps do not show hotspot areas nor recommend a safe route.
To ameliorate/minimize \emph{some of the above problems} and provide additional features,  our solution relies on location sharing via GPS/assisted GPS such as magnetic fingerprints and the BLE beacon RSSI (Received Signal Strength Identifier) nearby, for localization \cite{sanampudiindore} in an indoor setting. 
% Unlike existing solutions, our approach provides \emph{potential hotspot} information as well as \emph{ safe route} recommendation which is presented in sections \ref{hotspotdetection} and  \ref{saferoute} respectively. 

\section{Contact Tracing} \label{contactracing}
\begin{figure}
\centering
\includegraphics[width=5cm, height=5cm]{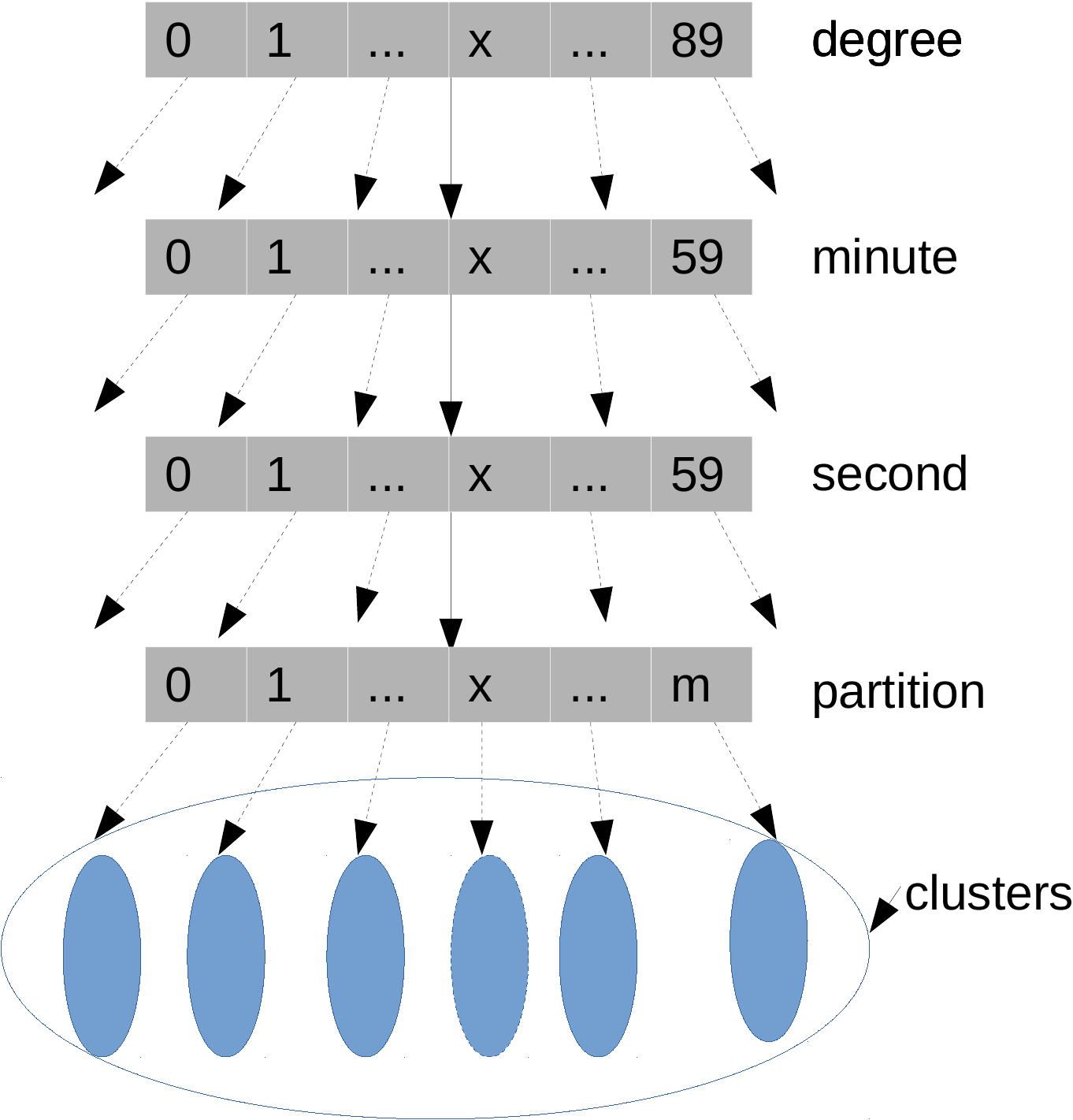}
\caption{M-way tree like data structure for storing coordinates }
\label{mwaytree}
\end{figure}

As discussed in the introduction section, our solution for contact tracing relies on the location sharing data, i.e., GPS location of the users collected through our app. The naive solution will be to compute the Euclidean distance (or \emph{Haversine distance} as shown in \eqref{haver}, where $\phi_i$ and $\lambda_i$ for $i \in \{1,2\}$ are lat/long of two users,
 if one wants to be more exact) every $t$ mins for finding the contact event. 
 \begin{equation}\label{haver}
     {\scriptstyle  d = 2r \arcsin\left(\sqrt{\sin^2\left(\frac{\phi_2 - \phi_1}{2}\right) + \cos(\phi_1) \cos(\phi_2)\sin^2 \left(\frac{\lambda_2 - \lambda_1}{2}\right)}\right)}
\end{equation}
 A contact event is defined as when two users are within a distance of $d$ meter for at least $t$ mins. The naive solution for contact tracing requires computing pairwise distances and has the complexity of $O(n^2)$ for $n$ users and hence not feasible for large $n$. Therefore, we discuss a data structure which is  M-way \emph{like} tree data structure for storing GPS locations (latitude/longitude) as shown in fig. \ref{mwaytree}. In other words, we map latitude/longitudes to the proposed tree as discussed later in details. Further, latitude and longitudes are mapped to two separate trees for parallel computation of the contact distances. How do we arrive to the direct distance between two users from latitude and longitude distances is presented in detail in section \ref{intuition}. Note that the M-way like tree is not the same as k-d tree which is binary tree and requires comparison for insertion. Whereas M-way like tree  involves look-up operation and no comparison. In this spirit, it is a kind of multi-level hashing.

A GPS coordinate which consists of latitude and longitude is usually expressed in the degree, minute, and second format (DMS), e.g., $(28\degree50' 30.12462")$. For latitude and longitude, degree ranges in $[0,1,\ldots,89]$ and $[0,1,\ldots,179]$ respectively while minutes and seconds vary in $[0,1,\ldots,59]$ for both which are shown in the top 3 gray levels in fig. \ref{mwaytree}. For notational brevity, we do not show degree ($\degree$), minute (') and seconds (``) symbols in the fig. The last gray level is the partition of the seconds into $m$ equal parts.   As the most interesting property of M-way like tree data structure  is the partition at the leaf, i.e., partition every {\it second} of GPS (which is equal to approx. 30 meters) to $m$ equal parts so that $m = 30/d$. Here, $d$ is our contact distance which is a tunable parameter ($d=5$ in our case). The benefit of the partition is that all users within a distance of $d$ meters on latitude will fall into the same leaf (bucket) and form natural clusters. For example, consider $d=5$ and hence $m=6$, then the intervals for the fractional part of the seconds (which varies in [0,1]) at the leaf node will be $[0,1/6),[1/6,2/6),\ldots,[5/6,1]$. Users $u_1$ and $u_2$ whose latitudes are $(28\degree50'30.12462")$ and $(28\degree50'30.05462")$ respectively will map to the same leaf node because their fractional part of the second (.12462 and .05462) will map to the same leaf and thus are in the same cluster.
\begin{figure}
\centering
\includegraphics[width=6cm]{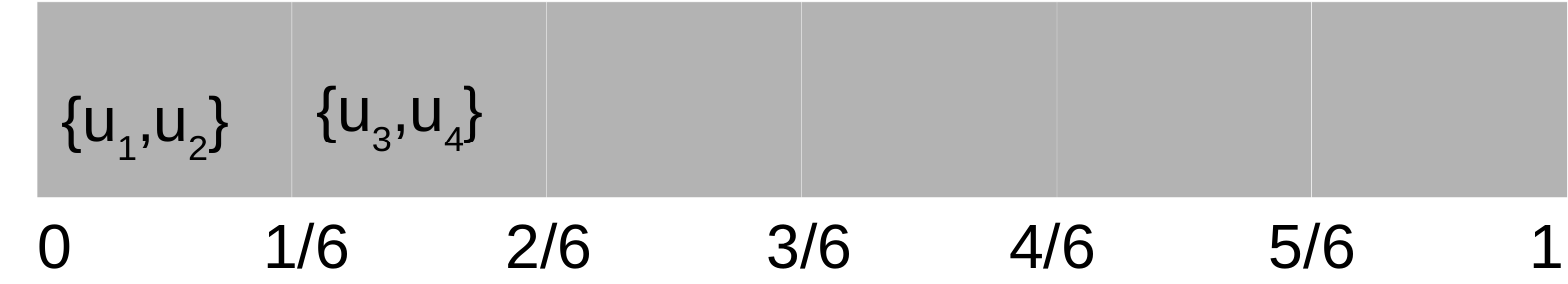}
\caption{Example to handle cases when users in the neighboring clusters may be within contact event distance $d$ and not captured during mapping of lat/long to the M-way like tree.}
\label{boundarycase}
\end{figure}
The approach presented above misses users who are within contact event distance $d$ but fall in the neighboring clusters.  For example, users $u_1$ and $u_3$ whose latitudes are $(28\degree50'30.12462")$ and $(28\degree50'30.17462")$ will map to different clusters but are within the contact event distance $d=5m$. An example of such a case is shown in fig. \ref{boundarycase} with four users. Out of these four users, two fall in one cluster, and the other two fall in a different neighboring cluster. To capture missed cases during the mapping of lat/long to the tree, we perform pairwise distance computation. For example, we need 4 distances computed ($\{u_1,u_3\},\{u_1,u_4\},\{u_2,u_3\},\{u_2,u_4\}$) in the worst case (worst case occurs when half of the users fall in one cluster and the other half fall in the neighboring cluster). Had we not mapped users to the tree,  ${\binom{n}{2}}$ computations are required with quadratic complexity $O(n^2)$ and not preferred for large $n$. To see this in fig. \ref{boundarycase}, instead of 6 pairwise distance computations in our running example, we needed only 4 pairwise distance computations in the worst case and hence saving two distance computations. Such a saving becomes paramount for large $n$.

We collect the GPS locations using our in-house app and send it to the cloud every $t/2$ mins for tracking users at least for $t$ mins ($t=5$ mins in our experiments). That means we map GPS locations to M-way like tree structure every $t/2$ mins. We have two trees for every interval of $t/2$ mins: one at time step $t/2$ and the other at $t$. At this step, we perform the intersection of leaves, and the ID of the users found in the intersection is stored. This method guarantees that two users in the same cluster have been in contact for at least $t/2$ mins. It also ensures that it will capture all users who have spent at least $t$ mins together within $d$ meter distance. One limitation is that it might miss users who have been in contact for 3 or 4 mins. Since our goal is to find all users who came in contact for at least 5 mins or more, we can ignore such corner cases.
\begin{figure}
\centering
\includegraphics[width=8cm]{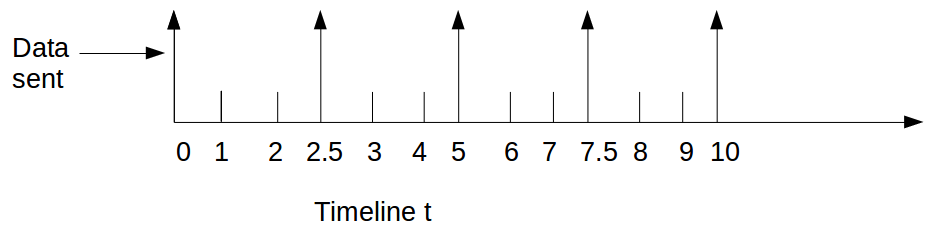}
\caption{Timeline to show when to send GPS coordinates to the cloud}
\label{timeline}
\end{figure}
\subsection{\bf Intuition behind \texorpdfstring{$t/2$}{Lg} mins}
It is not necessary to send GPS coordinates every $t/2$ mins. However, to guarantee that two users make contact for at least $t$ mins for the infection to occur, it is necessary to send GPS data every $t/2$ mins as shown in fig.  \ref{timeline}  for $t=5$ mins. 
We can see in the fig. \ref{timeline} that if two users met at time $t_1=1$ and remain in contact until $t_2=6$ ($\Delta t = t_2-t_1$) then we can see that we have sent their data two times (at $t=2.5$ and $t=5$ mins). As a result, we can capture the contact event. On the other hand, if we are sending data every $5$ mins, it is not possible to capture the aforementioned contact event. This approach guarantees that all users whose contact event duration is at least $5$ mins will be definitely captured. However, it may miss contact events of duration 3 or 4 mins. 
\subsection{\bf How lat/long distances map to circular \texorpdfstring{$d$}{Lg} m?}  \label{intuition}
\begin{figure}
\centering
\includegraphics[width=4cm, height=4cm]{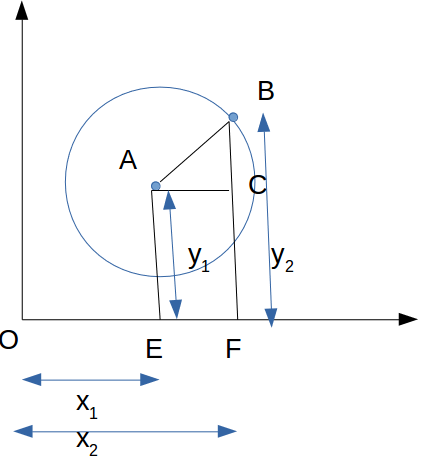}
\caption{Mapping lat/long to circular distances.}
\label{circular}
\end{figure}

We want to compute the distance $d = AB$ as shown in the fig. \ref{circular}. That is, all users at a distance $d$ from user $A$. For small distances ($d<10$ or 15 m in our case), we can safely assume that the points $A$ and $B$ lie on Euclidean coordinate system. Therefore, we can draw a simple circle at $A$ whose radius is $d$ (otherwise, it is {\bf great circle}). 

{\bf Main challenge:} computing $d$ is easy (we can take Euclidean distance or Haversine to be more precise). But, Euclidean computing distance for a large number of users (n) is computationally challenging. It takes $O(n^2)$ for pairwise distances. When $n=1$M, it takes $10^{12}$ distances to compute and storing them requires $O(n^2)$ space. Thus, pairwise Euclidean distance computation is impractical since we want a scalable solution where $n$ can be as large as 1.3B. Therefore, our main idea is to avoid computing $d$ directly and instead compute $AC$ and $BC$ distances, which are lat/long distances between $A$ and $B$. 
\begin{figure}[t]
\centering
\includegraphics[width=2.5cm, height=3cm]{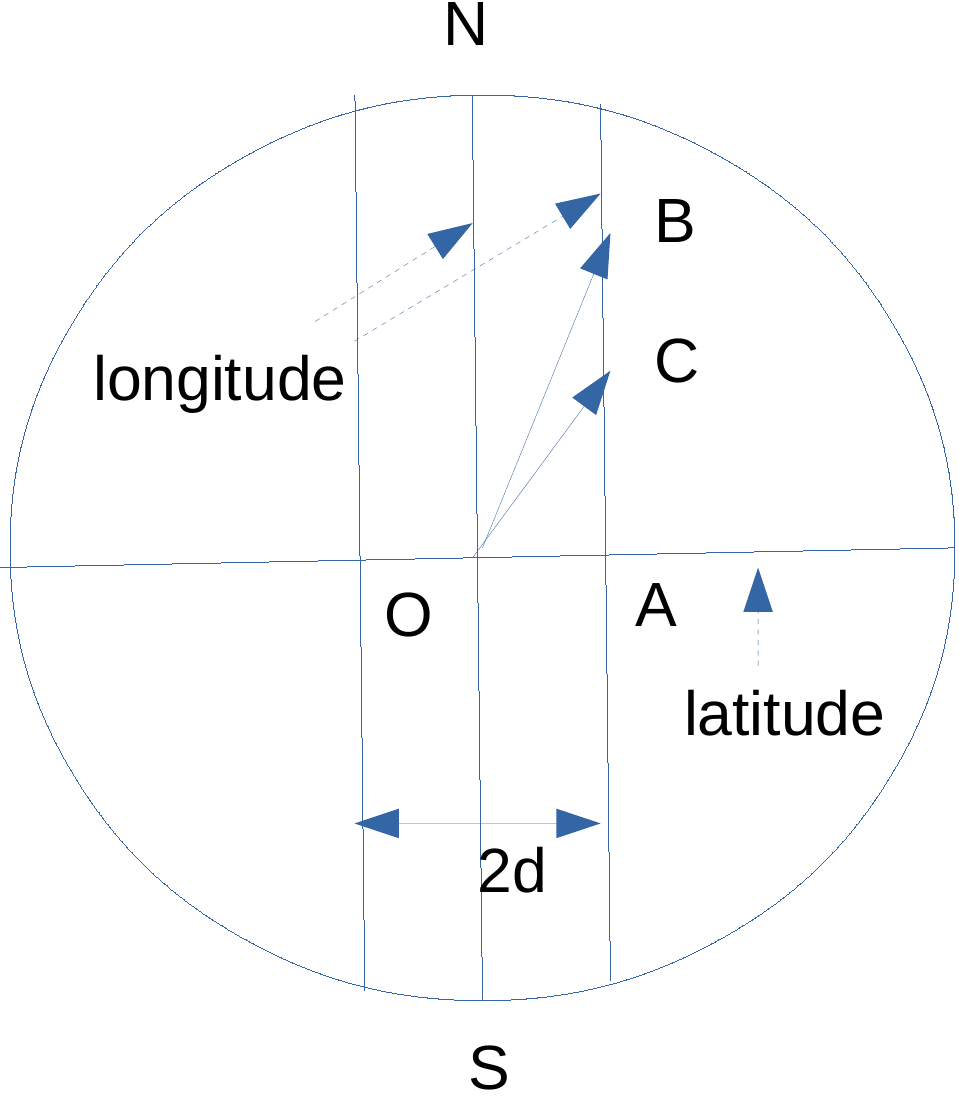}
\caption{Figure showing the case when many users are within contact distance $d$ along latitude but can be for away along longitude.}
\label{same_lat}
\end{figure}

Pythagorean theorem says that $AB^2 = AC^2 + BC^2$. Thus, if we can compute lat/long distances, we can find $d$. For $d = 5$m. We can find either (lat, long)=(3m,4m) or (4m, 3m). That means we find all users who are 3m away from user in question on latitude and 4m on longitude or vice-versa. Choosing (lat, long)=(3m,4m) gives a more accurate solution since computing distance along longitude depends on the accuracy of the distance along latitude \cite{wiki:xxx}. Further, notice that we map latitude/longitude of a user to two separate M-way like trees. That means we create a tree for mapping latitude and a tree for longitude. Such trees are created every $t/2$ mins, and users making contact events are found by the intersection of leaf nodes. More details of the procedure is described in the next section.
Approximating diagonal distance via distances along latitude/longitude does not always hold. In the previous example, when $(lat, long)=(3m,4m)$, we are finding users who are within 3m (4m) along latitude (longitude) and clustering them through mapping their latitude (longitude) on the M-way like tree data structure. However, the catch here is that there is a possibility that users within the contact distance along latitude but far away along longitude (or vice versa) are mapped to the same leaf node. As shown in fig. \ref{same_lat}, users at locations B and C will fall in the same leaf node when we map longitudes to the tree; however, locations B and C can be miles away along latitude (parallel lines passing through B and C but not shown in the figure). This can degenerate into a case when many users are mapped to the same leaf node of either tree. Such a situation can be avoided by running our proposed methodology over each city instead of the whole country for contact tracing. What will happen is that the probability that many users align to the same latitude (longitude) simultaneously for a short duration will be very low due to population dynamics. Therefore, we assume that a leaf node will not have many users mapped to it in the worst case.
In the next section, we present solutions when the users stay at the same location for at least $t$ mins (called the static case) and moving (called dynamics case). Though dynamic case subsumes the static case, we bifurcate the mapping into two cases for speed-up purposes.
\subsection{\bf Static Case}
\begin{figure}
\centering
\includegraphics[width=3.5cm, height=3cm]{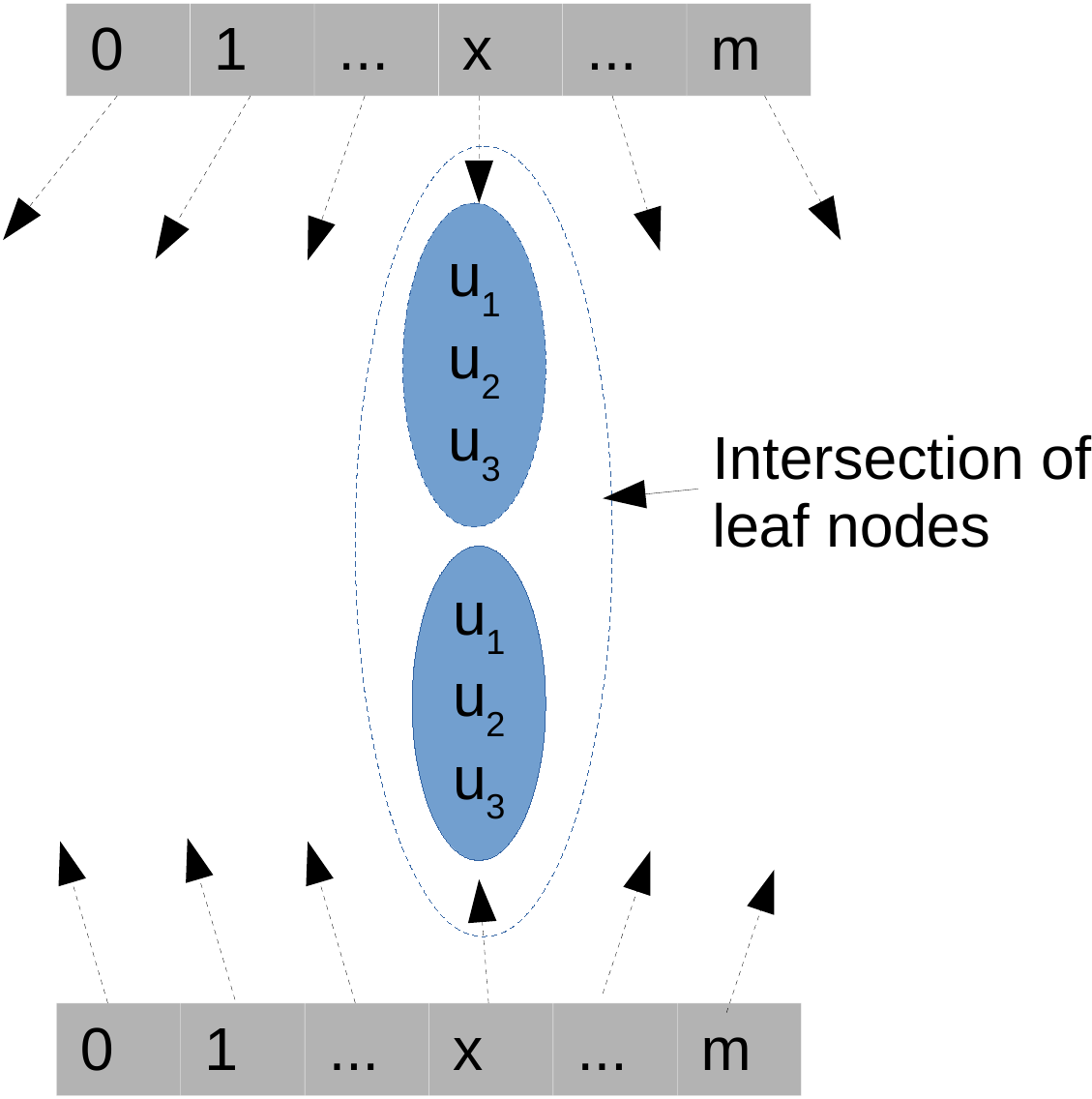}
\caption{Finding users in close proximity through intersection of corresponding leaves  of two M-way like tree in the static case}
\label{static}
\end{figure}

A case is static when users who participate in a contact event are not moving at least for $t$ mins. A static case can be easily handled by mapping lat/long to the M-way like tree every $t/2$ mins. The intuition is that two users who stay within a distance of $d$ m for at least $t$ mins will map to the same leaf node of the M-way like tree. We then find the intersection of the two M-way like trees so obtained (intersection of two corresponding leaves to be more precise since users are static). Users in the intersection will be our output candidates (see fig. \ref{static} where leaves from two trees are shown facing each other).
Illustration of the static case: In the fig \ref{static}, lat/longs are mapped to two trees as discussed before. The user $u_1, u_2$ and $u_3$ were at a location at time $t$ (top leaf node). After $t+\Delta t$ ($\Delta t=2.5$ mins), users $u_1, u_3$ and $u_3$ were found at the same location resulting in the leaf node shown in the bottom. When the intersection is performed, $u_1, u_2$ and $u_3$ are found to be present at the same location. That means, user $u_1,u_2$ and $u_3$ participated in a contact-event. The entire process of the static case is described step by step in Algorithm \ref{alg:static}.
\begin{algorithm}
\DontPrintSemicolon
\KwIn{$T_t$ - M-way tree at time $t$, $T_{t+1}$ - M-way tree at time $t+\Delta t$}
\KwOut{CT - Contact Tracing Vector}
\For{$d$ in $degrees$} {
\For{$m$ in $minutes$} {
\For{$s$ in $seconds$} {
\For{$p$ in $partitions$} {
  $ct=T_{t+1}[d][m][s][p] \cap T_t[d][m][s][p]$\;
  $\textit{Insert ct in CT}$\;
  \For{$n$ in [$p-1$, $p+1$]}{
  \If{$T_{t}[d][m][s][n]$ and $T_{t}[d][m][s][p]$ has peoples who are less than 5mt apart}{
  $end\_case_{t}$ = $findpairs($
  $T_{t}[d][m][s][n],T_{t}[d][m][s][p])$\;
  }
  \If{$T_{t+1}[d][m][s][n]$ and $T_{t+1}[d][m][s][p]$ has peoples who are less than 5mt apart}{
  $end\_case_{t+1}$ = $findpairs($
  $T_{t+1}[d][m][s][n], T_{t+1}[d][m][s][p])$\;
  }
  }
  $Insert$ $end\_case_{t} \cap end\_case_{t+1}$ in $CT$\;
  }}}
}
\Return{$CT$}\;
\caption{Static Case. [] denotes indexing in an array/dictionary}
\label{alg:static}
\end{algorithm}
\subsection{\bf Dynamic Case}
\begin{figure}
\centering
\includegraphics[width=5cm, height=4cm]{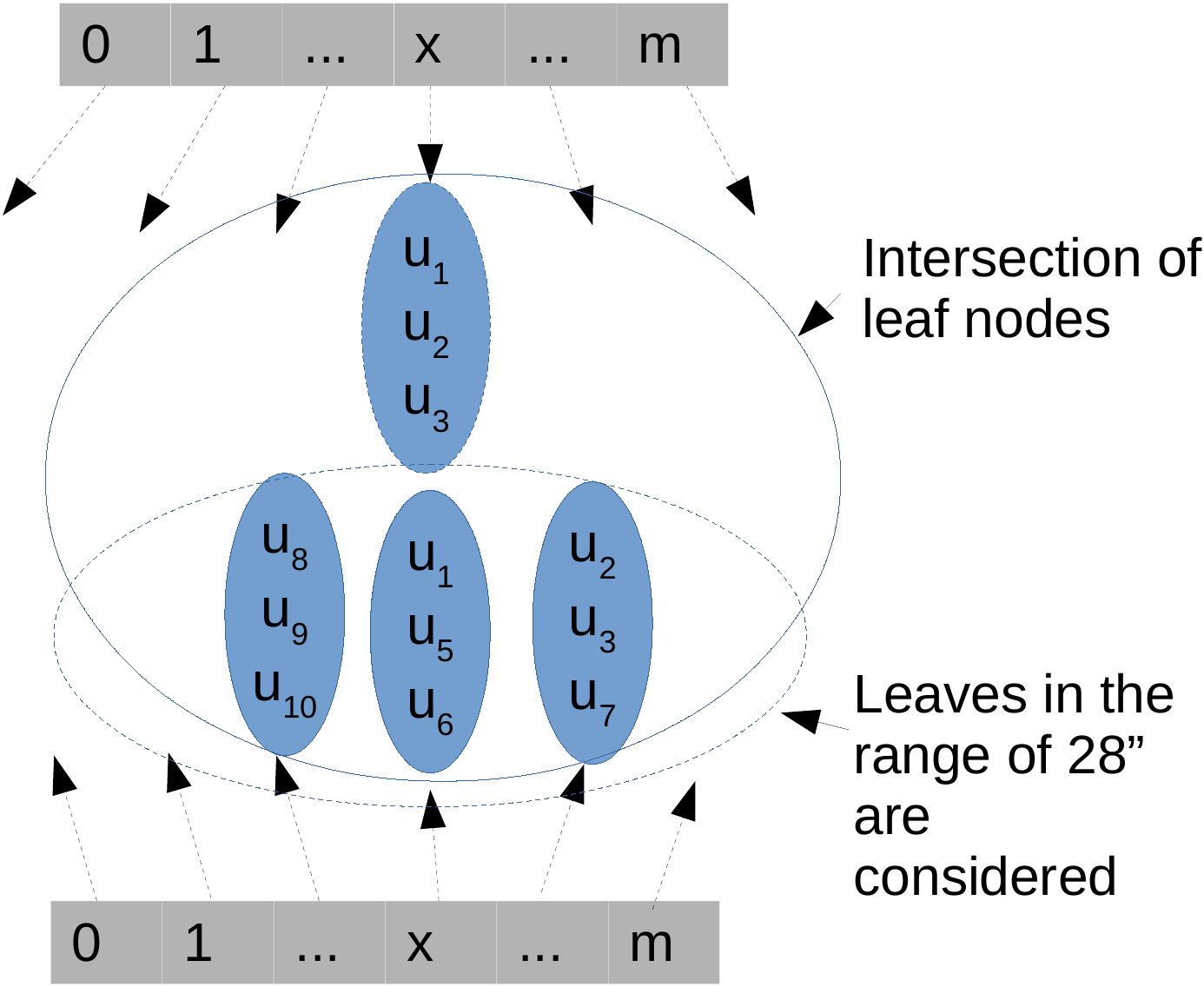}
\caption{Finding users in close proximity through intersection of corresponding leaves  of two M-way like tree in the dynamic case}
\label{dynamic}
\end{figure}

The dynamic case is when users move together in close proximity (within a radius of $d$ m). Due to the movement, their lat/log are continuously changing, resulting in their mapping to two different (non-corresponding) leaf nodes in the M-way like tree. To find out where users in close contact map to the leaf require performing the intersection of each leaf node against all nodes in the latter tree. This is computationally intractable, requiring $O(n^2)$ intersection. Therefore, we solve this issue via an approximation. First, we discuss when users are walking on foot (movement in a car can be handled similarly). On avg, a pedestrian speed is 1.4 m/s. That means, in 5 mins, they  walk around 420 meters which maps to 14” in lat/long. Therefore, we perform the intersection of each leaf against all leaves in the latter tree which are 14” left and right of the corresponding leaf as shown in the fig. \ref{dynamic} (left/right is considered since users can move left side or right side of the current latitude. Similarly, up/down movement might happen for longitude).
In either case (static/dynamic), once we find users who made contact along the latitude and longitude, we need to compute the distance AB (the actual contact direction) as discussed in section \ref{intuition}. To build the final contact list, we proceed as follows. Firstly, we create a set of pairwise contacts from the contact list found by the intersection operation as mentioned previously. This step is repeated for the longitude contact list as well. Secondly, the two sets are intersected to generate the final contact list, i.e., the list of users who actually made a contact event along the direction AB. This operation cost $O(n)$ (due to sets being a hashmap) where the hidden constant is the length of the longest contact list, which is bounded above by a constant since a user can not meet more than a certain number of users in $t$ (=5) mins. From the contact list, we can easily create a {\bf contact vector} of each user and store in an adjacency matrix as discussed in the next section. The entire process of the dynamic case is described step by step in Algorithm \ref{alg:dynamic}.

\begin{algorithm}
\DontPrintSemicolon
\KwIn{$T_t$ - M-way tree at time $t$, $T_{t+1}$ - M-way tree at time $t+\Delta t$}
\KwOut{CT - Contact Tracing Vector}
$q = 14$ \hspace{2cm} \# 420 meter = 14"\;
\For{$d$ in $degrees$} {
\For{$m$ in $minutes$} {
\For{$s$ in $seconds$} {
\For{$p$ in $partitions$} {
$p1 \gets []$\;
$p2 \gets []$\;
  \For{$n$ in [$p-q$, $p+q$]}{
  $ct=T_{t+1}[d][m][s][n] \cap T_t[d][m][s][p]$\;
  $\textit{Insert ct in CT}$\;
  \For{$n1$ in [$n-1$, $n+1$]}{
  $end\_case_{t}$ = find peoples who are less than 5mt apart in $T_{t}[d][m][s][n1]$\;
  
  $end\_case_{t+1}$ = find peoples who are less than 5mt apart in $T_{t+1}[d][m][s][n1]$\;
 
  }
  $Insert$ $end\_case_{t}$ in $p1$\;
  $Insert$ $end\_case_{t+1}$ in $p2$\;
  }
  $Insert$ $p1 \cap p2$ in $CT$\;
  }
  }}
}
\Return{$CT$}\;
\caption{Dynamic Case. [] denotes indexing in an array/dictionary}
\label{alg:dynamic}
\end{algorithm}

\section{Potential Hotspot Detection} \label{hotspotdetection}
Our methodology can be used to detect \emph{potential} hotspot areas in a city. Note that there is no unanimous definition of what constitutes a hotspot in epidemiology. As per \cite{lessler2017hotspot}, hotspot can be defined in three ways: (a) as areas of disease elevated occurrence or risk, (b) as areas of frequent disease emergence or reemergence, and (c) an area of elevated transmission efficiency. We take the third definition that is based on the assumption that large gatherings can stimulate the rate of transmission. Therefore, we identify areas in a city where large gatherings are happening along with  COVID +ve cases and predict those areas as potential hotspots. The potential hotspot information can help safeguard users and discourage them from freely moving in those areas. Note that our method does declare gatherings as \emph{potential hotspot} if there are no +ve cases in that gathering (cluster). Our assumption is that either COVID +ve user is surrounded by normal users who are not quarantined, and in some cases where COVID +ve user might also be moving, say coming back from a testing center. So, hotspot may contain COVID +ve users who are moving.

\begin{figure}
\centering
\includegraphics[width=5cm, height=1cm]{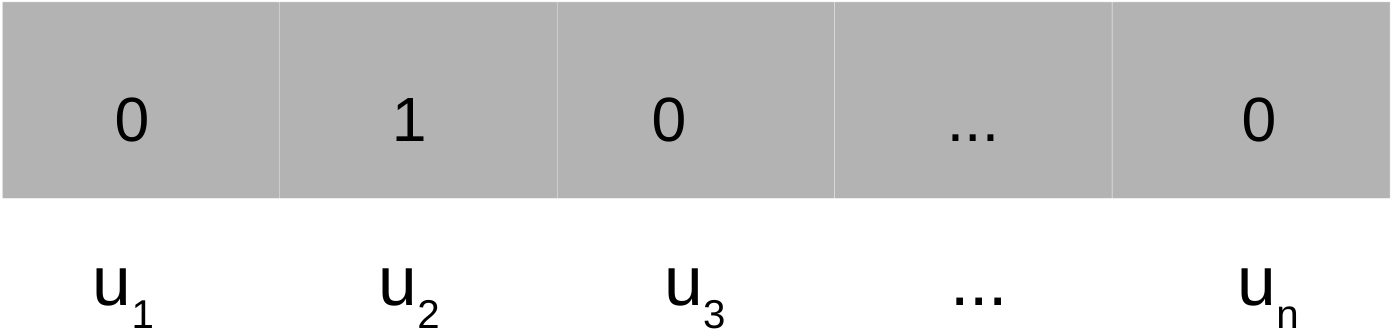}
\caption{Bit vector to store COVID +ve users}
\label{bitvec}
\end{figure}

 In order to locate areas of a potential hotspot, we proceed as follows. Contact list of each user obtained from the contact tracing methodology presented in section \ref{contactracing} from the last 14 days is stored in the form of an adjacency matrix as shown in \eqref{adlist}.
\begin{equation} \label{adlist}
\begin{blockarray}{ccccc}
& u_1 & u_2  & \dots & u_n \\
\begin{block}{c[cccc]}
 u_1 & 1  & 1 & \dots & 1  \\
 u_2 & 0  & 0 & \dots & 1  \\
  \vdots & \vdots  & \vdots & \ddots & \vdots  \\
  u_n &1 & 0  & \dots & 1  \\
\end{block}
\end{blockarray}
 \end{equation}
(In actual implementation, it can be implemented using a hashmap where keys are user ids and values are queue/list which stores the ids of users in the order the contact happens). We call each row as a {\bf contact vector} for that user. We can easily maintain the contact vector by adding /deleting from the queue/list the new/old contacts. Note that the adjacency matrix stores users who came in contact with other users. It does not store the COVID +ve users. In other words, we rely on official data from government authority to declare a user as a COVID +ve based on the diagnostic report. Once we receive such information, we can mark that user as a COVID +ve by setting a bit along that row or by maintaining another bit vector which is indexed by the user id as shown in fig. \ref{bitvec}.

To detect a \emph{potential hotspot} area, we provide a reference GPS location. This can be a city location as searched by users interested in knowing hotspot areas. The reference GPS is mapped to the lat/long tree created in that period. Assuming a 10 km radius from the reference point though, this can be dynamically adjusted according to the city area. Firstly, all the users are identified who are within the 10 km radius from the reference point. The 10 km of distance maps to around 5.39 minutes in lat/long. Therefore, all the users in the leaves, which are 5.39' left and right of the reference point, are identified. The user ids of such users are mapped to the bit vector (fig. \ref{bitvec}) to find out users who are COVID +ve within our search area (10 km, for example). Using the adjacency matrix, we also identify users who came in contact with users who are found COVID +ve in the search area. Such users might form \emph{susceptible users}.
Further, user ids of the susceptible users are cross-tallied in the search area (since users who came in contact with the COVID +ve user in the search area $t$ days ago may have moved to a different city). Finally, we have built a set of COVID +ve users and suspected users. We can mark COVID +ve and susceptible users on the city map, and if the number of COVID +ve users exceeds a threshold, it can be marked as a \emph{potential hotspot city}.
\section{Safe Route Recommendation}\label{saferoute}
Based on the potential hotspot areas, we can recommend a safe route for travel. Users currently use google map/Apple Maps as a primary navigation medium for traveling from one place to another. However, these maps have limited information about whether a route is safe for travel. For example, unless someone marks specifically that a particular route is blocked or traffic is slow, current navigation systems have limited information to tell users in advance that route is not safe. This situation was observed during the recent exodus of people from red zones such as Mumbai/New Delhi. People were following google maps to travel to their hometowns. However, the map took them to other hotspot cities on the way. As a result, they have to take a long detour once they reached the hotspot city on the way because entry into the city was not allowed. In such situations and many others, such as avoiding potential gatherings that may be a potential hotspot, our approach can recommend a safe route, thereby avoiding hotspot areas in advance so that users can plan their travel accordingly.

\begin{figure}[b]
\centering
\includegraphics[width=3cm, height=2.5cm, scale=.3]{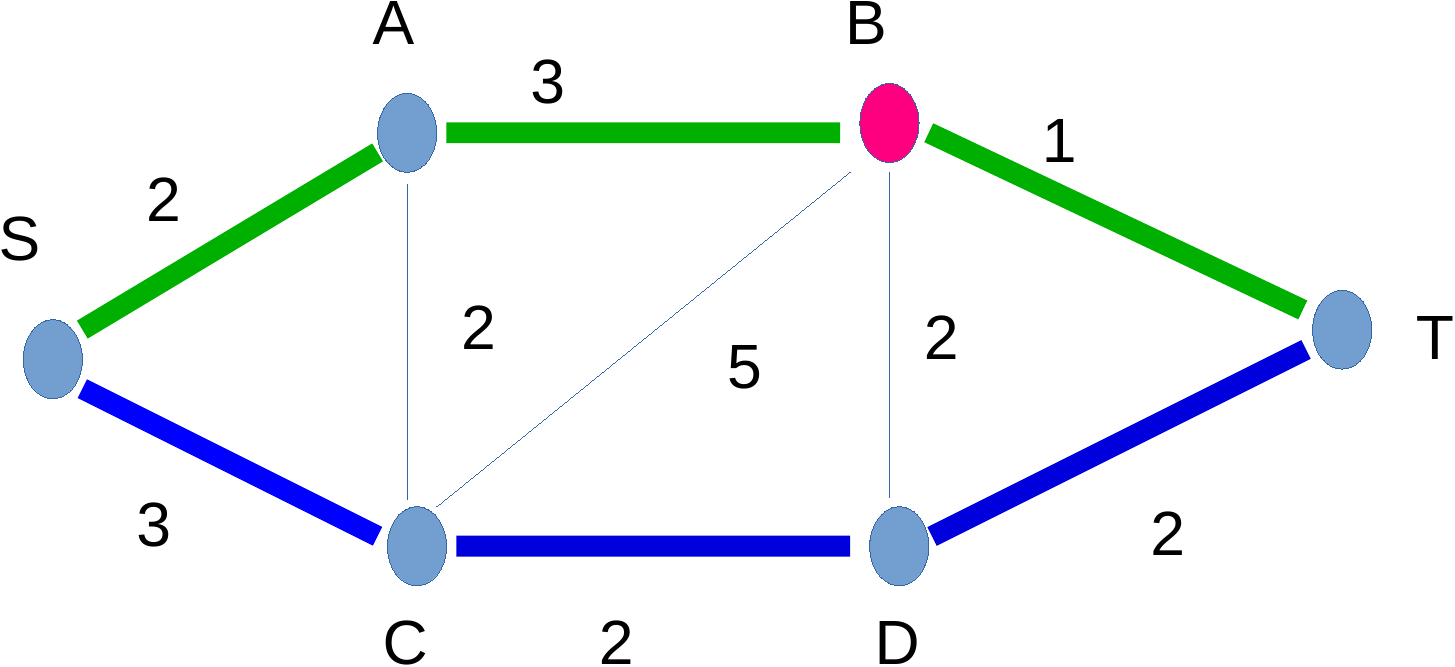}
\caption{Safe route recommendation. The shortest route might not be safe. Google Maps recommend the Green route since it is the shortest. Our approach recommends the Blue route though it may not be the shortest, it avoids the potential hotspot areas.}
\label{hotspot}
\end{figure}

Suppose a user wants to travel from source location $s$ to target location $t$ as shown in fig. \ref{hotspot}. If the user follows Google map, it might show the {\it green} route since it is the shortest path. However, it may take the user to the hotspot city since it does not have {\it potential hotspots} information in {\it real-time}. Following our previous approach, we have the information of {\it potential hotspots}  based on the previous section's discussion.  Therefore, we remove the nodes representing the hotspot (red node in the fig. \ref{hotspot}) as well as edges incident to it and run the Dijkstra's algorithm \cite{VAUGHAN20209} \cite{dijkstra1959note} on the remaining subgraph. The algorithm produces the shortest path in the subgraph (marked in blue color). This path does not fall in red zones/hotspot areas and thus safe for travel. We verify our approach on a toy dataset and show its efficacy in the experiment section. The safe route recommendation algorithm is described in Algorithm \ref{alg:dijkstra}.
\begin{algorithm}[b]
\DontPrintSemicolon
\KwIn{$G$ - Graph, $S$ - Source, $D$ - Target, $HOTSPOT$ - list of hotspot nodes.}
\KwOut{Path from $S$ to $T$}
\For{each vertex $V$ in $G$} {
$distance[V] \gets \infty$\;
 $last[V] \gets None$\;
 \If{$V!=S$} {
 add $V$ to Priority Queue $Q$(Using Min-Heap)\;
 }}
 \While{$Q$ is Not Empty} {
 $U$ $\gets$ Extract min From $Q$\;
 \For{each unvisited neighbour $V$ of $U$}{
\If{$V$ not in $HOTSPOT$}{
$temporary\_Distance$ $\gets$ distance[$U$] + $edge\_Weight(U, V)$\;
\If{$temporary\_Distance < distance[V]$}{
$distance[V] \gets temporary\_Distance$\;
$last[V] \gets U$\;
}
}
 }
 }
\If{$distance[D]$ is $\infty$}{
\Return{$No Path$}\;
}
\Return{$distance[D]$}
\caption{Safe Route Recommendation Algorithm}
\label{alg:dijkstra}
\end{algorithm}
In the above algorithm, $G$ is the input graph, $S$ and $D$ are the sources and destination nodes (cities), and $HOTSPOT$ is the list of nodes marked as a hotspot thus not safe for travel. 
\section{ Complexity Analysis}
In this section, time and space complexity analysis of various operations is presented. 

{\bf Time complexity of the mapping} a GPS coordinate to a tree leaf is $O(h)$ where $h$ is the height of the tree. In our case, $h=4$, so mapping GPS coordinates to tree leaf is a constant time operation $O(1)$. Further, append operation of the users into the list attached to the leaf node will have a time complexity of $O(1)$.
After mapping, we need to find a new cluster of users missed in the mapping process as discussed in section \ref{intuition}.
Such users can be put into new clusters by sliding a window of size 2, moving over adjacent pairs of clusters, and computing Euclidean/Haversine distance for each pair of users from each cluster. Time taken by this operation is $O(ck\ell)$, where $k$ and $\ell$ are the sizes of two clusters and $c(=90*60*60*6)$ is a constant. As discussed previously, the size of the clusters can not grow unbounded in the worst case due to population dynamics. Usually, $k$ and $\ell$ are in the range of 1 to 100 since within a circle of 5m radius; we assume no more than  100 users can stay.   Further, the operation of finding missed users can be computed efficiently using parallelization (presented in the empirical section). 

{\bf Time complexity of the intersection} of leaf node (clusters) in the static case is more straightforward. We need to perform the intersection of $O(c)$ corresponding leaf nodes in two trees obtained at time intervals of $t/2$ mins. Here $c$ is as defined previously. Intersection operation in the dynamic case is involved and incurs extra overhead due to one leaf being intersected with 168 (=28"x6) leaf nodes in the other tree for the pedestrian case. Consequently, time cost goes up by a factor of 168 in the dynamic case compared to the static case. Further, note that the intersection operation in both the cases (static and dynamic) can be efficiently parallelized since one leaf node can be intersected with other nodes in parallel. Such parallelization tricks are used in the empirical section to reduce the runtime cost of the intersection operation. 

{\bf Space Complexity}:  Total number of internal nodes for mapping latitudes is $90 + 90*60+90*60*60 +90*60*60*m =329490+32400m$. Where $m$ is the number of partitions of a second. Take $m = 6$ for $d=5$,we have total internal nodes as 2273490. Assume each integer takes 8 bytes. The space complexity of an empty tree is approx 18.18 MB.
Further, we are just storing user ids, which for 10M users takes 80MB. In total, one tree after mapping 10M GPS coordinates will take 98.18MB of space. If we also store lat/long for 10M users, it will further take 160MB. Thus in the worst case, one tree will occupy approximately 260MB. (During execution, we store two trees every 2.5 mins interval. To generate contact vectors of 10M users takes around 17 mins as given in the experimental section; we maintain a total of 16 trees before we can flush the trees and reuse the space. Hence our approach takes roughly 4GB of space for generating contact vectors compared to 55GB in \cite{mahapatra2020dynamic}.)

 In our implementation, we take latitude $\simeq (7\degree,37\degree)$  and longitude $\simeq (68\degree,97\degree)$ for India for which empty tree takes  $\approx$39 MB when representing internal nodes as four-level dictionary keys and leaves as the empty list in the python programming language.  We store app IDs or some other useful global identifier such as IMEI number into leaf nodes to track each user. Thus space complexity for storing user id into leaf will be $O(n)$ where $n$ is the number of user ids. Therefore, the dominating part will be used for storing user ids, and it scales linearly with the users.

One crucial point to consider is that the leaf node is implemented as a list or array which has the cost of $O(1)$ for append operation. However, a growing list beyond specified size incurs extra overhead. However,  we must mention that \emph{any} list does not hold more than a constant number of users since each leaf maps GPS range of 5 meters and in a circle of 5-meter radius, we assume that not more than 100 users can stand.

\section{Empirical Demonstration}
In this section, we show the working of the proposed methodology for contact tracing, hotspot detection, and safe route recommendation on simulated data. 
\begin{table}[]
\caption{Run-time comparison in the static and dynamic cases. All time in seconds. Values in blue and black denote run-time in static and dynamic case respectively. Cells shaded in yellow color show serial run-time  whereas cells without shading indicate parallel run-time. Best viewed in color.}
\label{runtime}
\small
\begin{tabular}{|c|c|c|c|c|c|}
\hline
\multicolumn{4}{|c|}{Time to Map GPS to Tree}                                                                             & \multicolumn{2}{c|}{}                                                                         \\ \cline{1-4}
\multicolumn{2}{|c|}{Before(at T=t)}                        & \multicolumn{2}{c|}{After(at T=t+$\Delta$t)}                      & \multicolumn{2}{c|}{\multirow{-2}{*}{Intersection Tree Time}} \\ \hline
Latitude                     & Longitude                    & Latitude                     & Longitude                    & Latitude                                                      & Longitude                     \\ \hline
\rowcolor[HTML]{FFFFC7} 
{\color[HTML]{3166FF} 20.18} & {\color[HTML]{3166FF} 20.43} & {\color[HTML]{3166FF} 20.67} & {\color[HTML]{3166FF} 20.96} & {\color[HTML]{3166FF} 49.98}                                  & {\color[HTML]{3166FF} 46.66}  \\ \hline
\rowcolor[HTML]{FFFFC7} 
{\color[HTML]{000000} 21.99} & {\color[HTML]{000000} 22.09} & {\color[HTML]{000000} 23.45} & {\color[HTML]{000000} 23.67} & {\color[HTML]{000000} 150.89}                                 & {\color[HTML]{000000} 161.13} \\ \hline
\rowcolor[HTML]{FFFFC7} 
{\color[HTML]{3531FF} 30.43} & {\color[HTML]{3531FF} 32.98} & {\color[HTML]{3531FF} 35.87} & {\color[HTML]{3531FF} 34.88} & {\color[HTML]{3531FF} 113.67}                                 & {\color[HTML]{3531FF} 122.32} \\ \hline
\rowcolor[HTML]{FFFFC7} 
31.9                         & 33.88                        & 37.98                        & 34.98                        & 587.32                                                        & 632.49                        \\ \hline
\rowcolor[HTML]{FFFFC7} 
{\color[HTML]{3531FF} 46.76} & {\color[HTML]{3531FF} 48.87} & {\color[HTML]{3531FF} 43.42} & {\color[HTML]{3531FF} 49.44} & {\color[HTML]{3531FF} 496.75}                                 & {\color[HTML]{3531FF} 513.23} \\ \hline
\rowcolor[HTML]{FFFFC7} 
50.32                        & 47.64                        & 48,98                        & 51.32                        & 1208.98                                                       & 1267.22                       \\ \hline
\rowcolor[HTML]{FFFFC7} 
{\color[HTML]{3531FF} 55.43} & {\color[HTML]{3531FF} 51.4}  & {\color[HTML]{3531FF} 59.32} & {\color[HTML]{3531FF} 24.55} & {\color[HTML]{3531FF} 576.23}                                 & {\color[HTML]{3531FF} 612.43} \\ \hline
\rowcolor[HTML]{FFFFC7} 
65.12                        & 68.32                        & 66.23                        & 71.42                        & 2198.32                                                       & 2223.64                       \\ \hline
\rowcolor[HTML]{FFFFC7} 
{\color[HTML]{3531FF} 68.98} & {\color[HTML]{3531FF} 72.7}  & {\color[HTML]{3531FF} 74.89} & {\color[HTML]{3531FF} 79.84} & {\color[HTML]{3531FF} 786.67}                                 & {\color[HTML]{3531FF} 796.87} \\ \hline
\rowcolor[HTML]{FFFFC7} 
74.99                        & 76.65                        & 79.43                        & 80.1                         & 3983.78                                                       & 4073.27                       \\ \hline
{\color[HTML]{3531FF} 19.89} & {\color[HTML]{3531FF} 19.12} & {\color[HTML]{3531FF} 20.22} & {\color[HTML]{3531FF} 22.43} & {\color[HTML]{3531FF} 8.32}                                   & {\color[HTML]{3531FF} 7.39}   \\ \hline
22.43                        & 18.9                         & 21.32                        & 22.94                        & 22.69                                                         & 29.32                         \\ \hline
{\color[HTML]{3531FF} 30.21} & {\color[HTML]{3531FF} 33.41} & {\color[HTML]{3531FF} 33.78} & {\color[HTML]{3531FF} 36.98} & {\color[HTML]{3531FF} 32.43}                                  & {\color[HTML]{3531FF} 40.98}  \\ \hline
32.76                        & 36.74                        & 38.01                        & 39.63                        & 78.19                                                         & 102.23                        \\ \hline
{\color[HTML]{3531FF} 48.64} & {\color[HTML]{3531FF} 46.75} & {\color[HTML]{3531FF} 50.43} & {\color[HTML]{3531FF} 49.31} & {\color[HTML]{3531FF} 60.98}                                  & {\color[HTML]{3531FF} 73.92}  \\ \hline
50.76                        & 53.21                        & 48.98                        & 54.67                        & 365.42                                                        & 398.71                        \\ \hline
{\color[HTML]{3531FF} 59.45} & {\color[HTML]{3531FF} 62.32} & {\color[HTML]{3531FF} 58.43} & {\color[HTML]{3531FF} 56.49} & {\color[HTML]{3531FF} 86.15}                                  & {\color[HTML]{3531FF} 92.43}  \\ \hline
61.23                        & 65.43                        & 66.32                        & 68.32                        & 598.41                                                        & 631.42                        \\ \hline
{\color[HTML]{3531FF} 78.23} & {\color[HTML]{3531FF} 71.96} & {\color[HTML]{3531FF} 75.63} & {\color[HTML]{3531FF} 70.54} & {\color[HTML]{3531FF} 122.34}                                 & {\color[HTML]{3531FF} 134.43} \\ \hline
85.78                        & 79.12                        & 83.45                        & 81.12                        & 945.32                                                        & 990.87                        \\ \hline
\end{tabular}
\end{table}

\subsection{Contact Tracing Experiment}
To show the scalability of the contact tracing approach, uniformly at random GPS coordinates in the range latitude $\simeq (7\degree,37\degree)$  and longitude $\simeq (68\degree,97\degree)$ are generated for 2/4/6/8/10M users. We will map these GPS coordinates to the M-way tree like data structure as discussed and compute the contact vectors. Our goal is to estimate the time it takes to generate the contact lists. The serial and parallel versions for static and dynamic cases are implemented in C++ with openMP \cite{dagum1998openmp} (for multi-threading).  The experiments are conducted on a personal computer with Intel(R) Core(TM) \textit{i5-7500U} CPU and $8.0$ $GB$ memory (RAM) running Ubuntu operating system. The results are shown in the table \ref{runtime}. We have several observation from the results in the table. Firstly, mapping GPS to tree takes at most 80 secs for 10M users using serial implementation, whereas at most 86 secs in the parallel implementation. A little increase in time could be due to thread scheduling overhead in the latter case. In other words, mapping GPS to trees via parallel implementation is not \emph{recommended}, and serial implementation suffices to practical cases. Secondly, compared to mapping GPS coordinates to trees, performing intersection of leaves is time-consuming. Thirdly, intersection time using parallel implementation achieves dramatic speedup. For example, parallel implementation in the dynamic case for 10M users gains 4x speedup. Finally, generating the final contact vectors for 10M users takes around 16.5 mins (990.87+85.78 secs). In other words, if we want to track contact events every 5 mins, we need to maintain four sets of lat/long trees before we can flush the data in the trees and reuse them again, thereby reducing the memory footprint. 

\subsubsection{Baselines}
One of the baselines used to compare the runtime of the contact tracing approach is by performing pairwise comparison (the naive solution). The result from the serial and parallel implementation of the baseline is shown in Table \ref{baselin}. It is obvious that our approach beats the naive solution by a significant order of magnitude and favors practical implementation.
% \usepackage{multirow}
\begin{table}[]
\small
\centering
\caption{The runtime comparison of our approach with the baseline. The experiment is conducted over 2M users. The number of threads uses in the parallel implementation is set to 10.}
\label{baselin}
\begin{tabular}{|l|l|l|}
\hline
                                                 & Implementation type & CPU time in seconds \\ \hline
\multicolumn{1}{|c|}{\multirow{2}{*}{Baselines}} & Serial              & 1135714.0        \\ \cline{2-3} 
\multicolumn{1}{|c|}{}                           & Parallel            &   127124.59              \\ \hline
\multirow{2}{*}{Our approach}                    & Serial              & {\bf 185.80}          \\ \cline{2-3} 
                                                 & Parallel            & {\bf 52.26}           \\ \hline
\end{tabular}
\end{table}
\subsection{Hotspot Detection Experiment}
\begin{figure}
    \centering
  \subfloat[At time $t_{0}$]{\includegraphics[height=4cm]{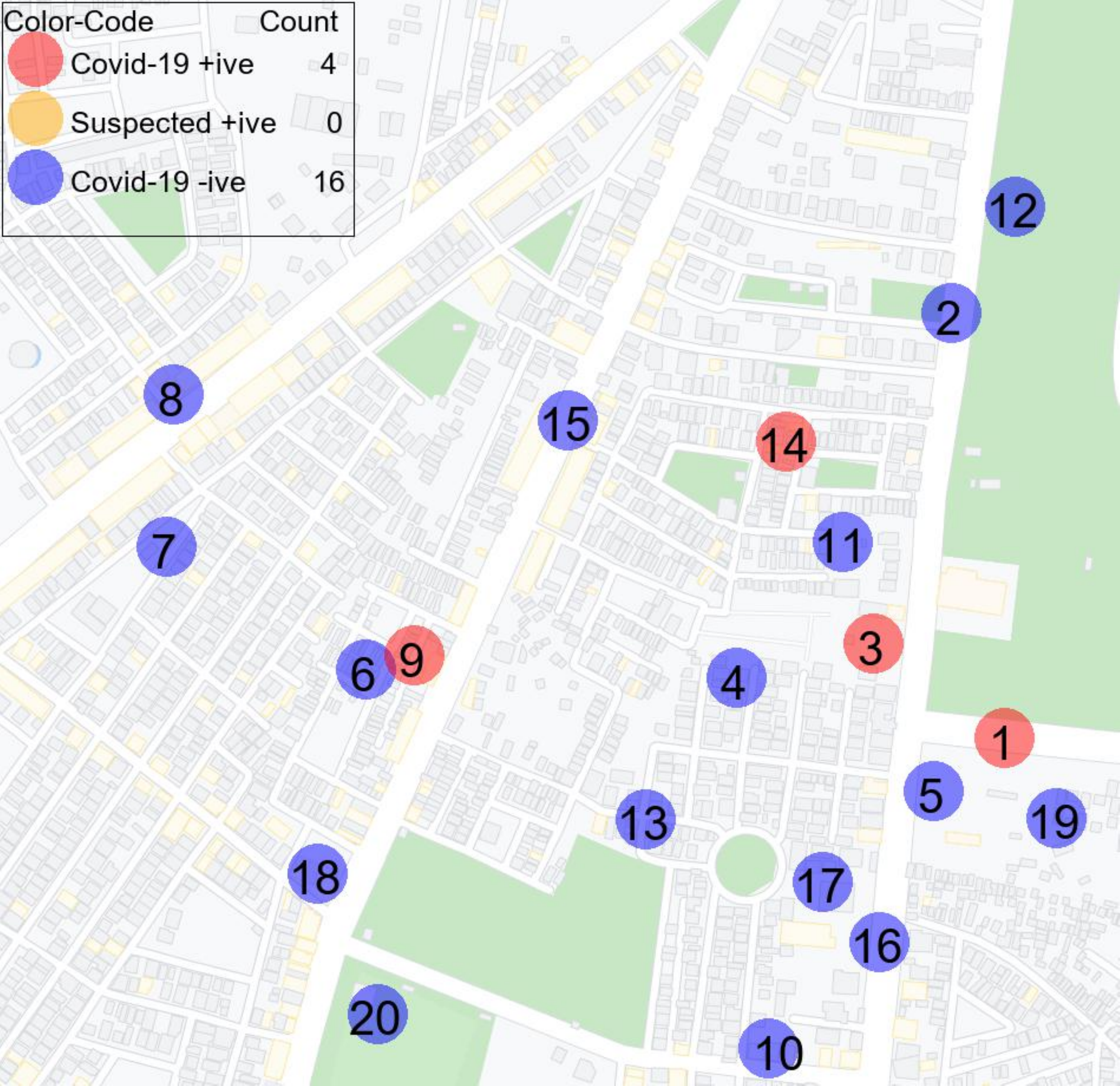}}
    \hfill
    \subfloat[At time $t_{1}$]{\includegraphics[height=4cm]{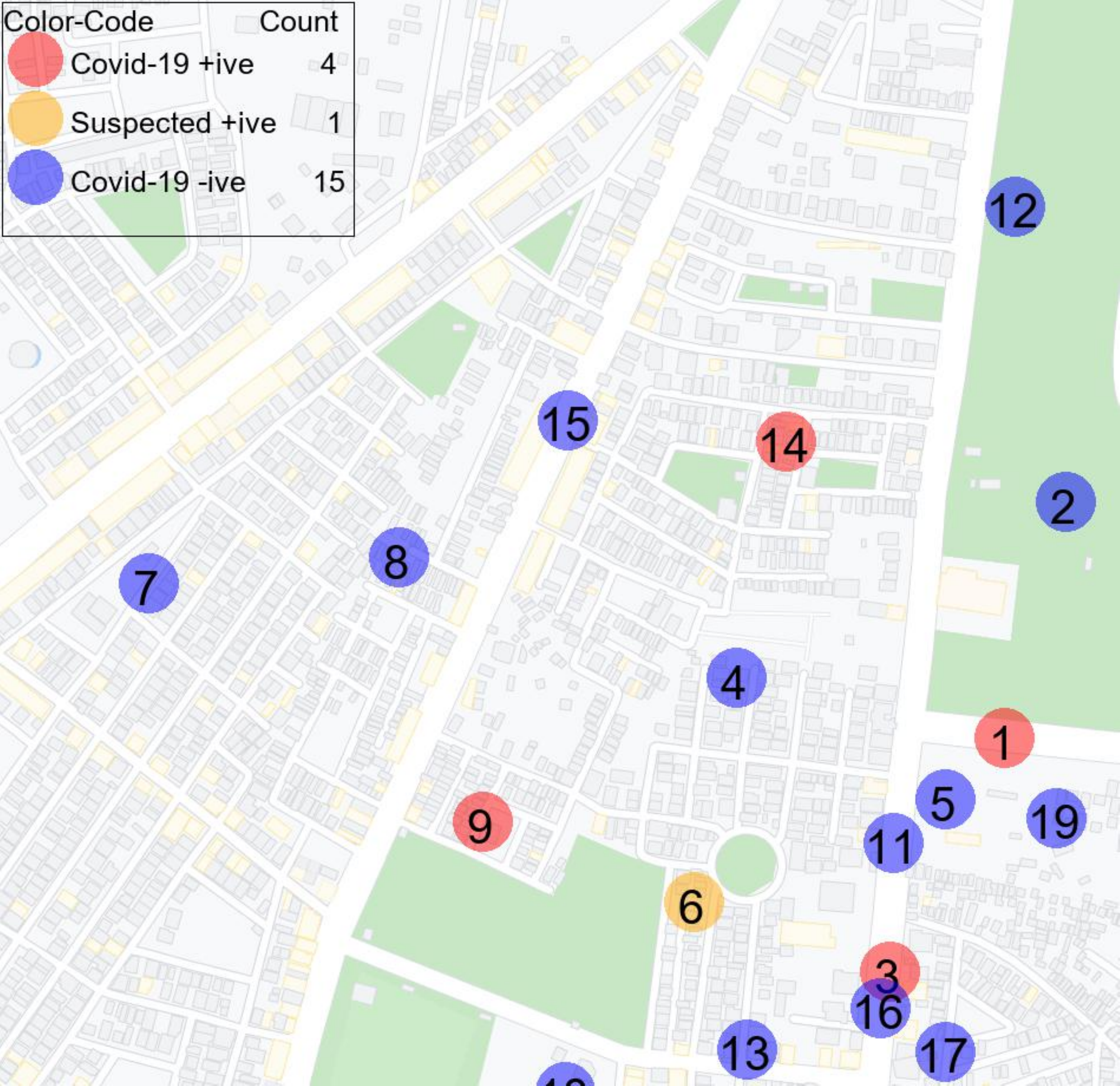}}
    \\
    \subfloat[At time $t_{2}$]{\includegraphics[height=4cm]{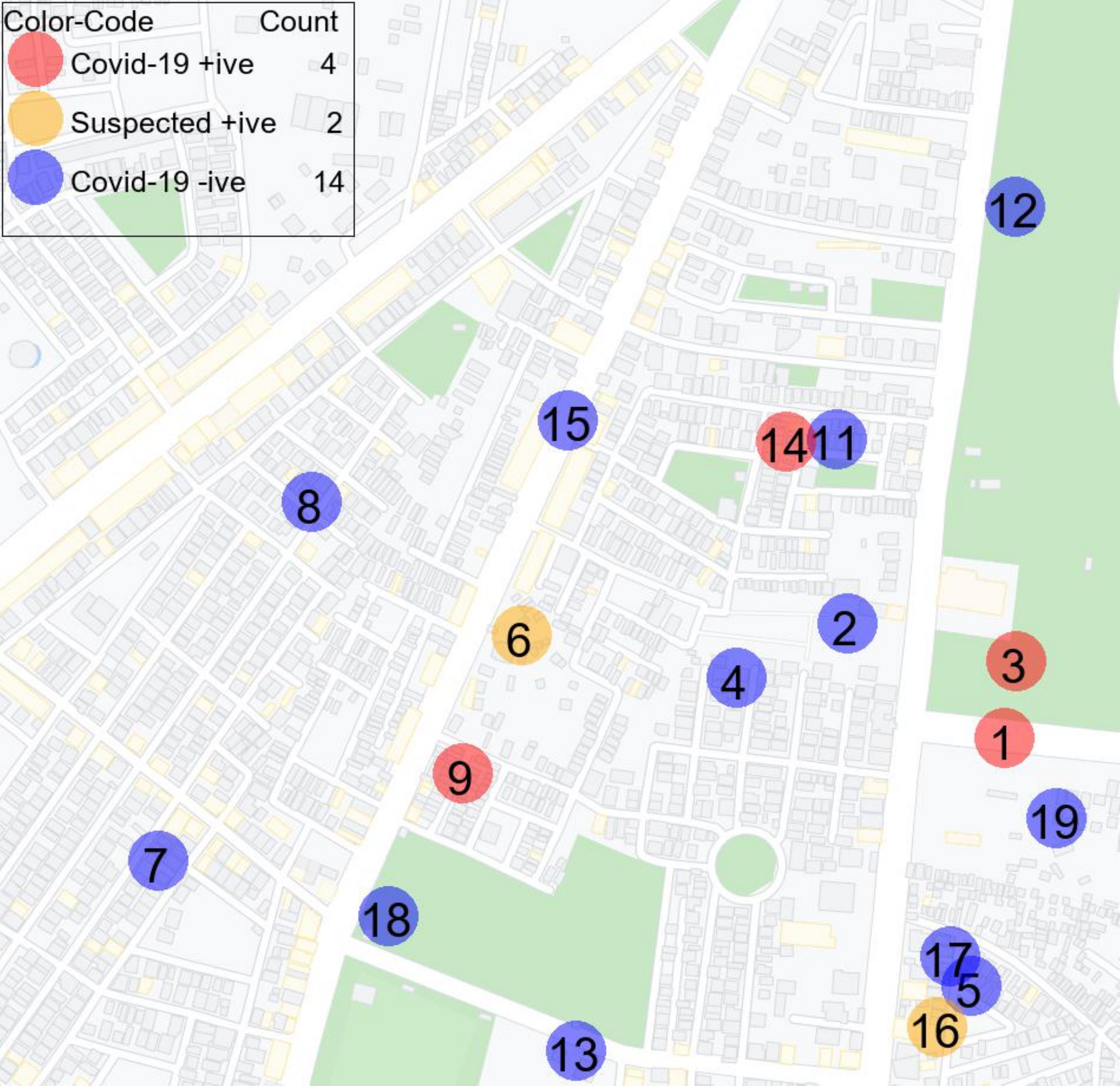}}
    \hfill
    \subfloat[At time $t_{3}$]{\includegraphics[height=4cm]{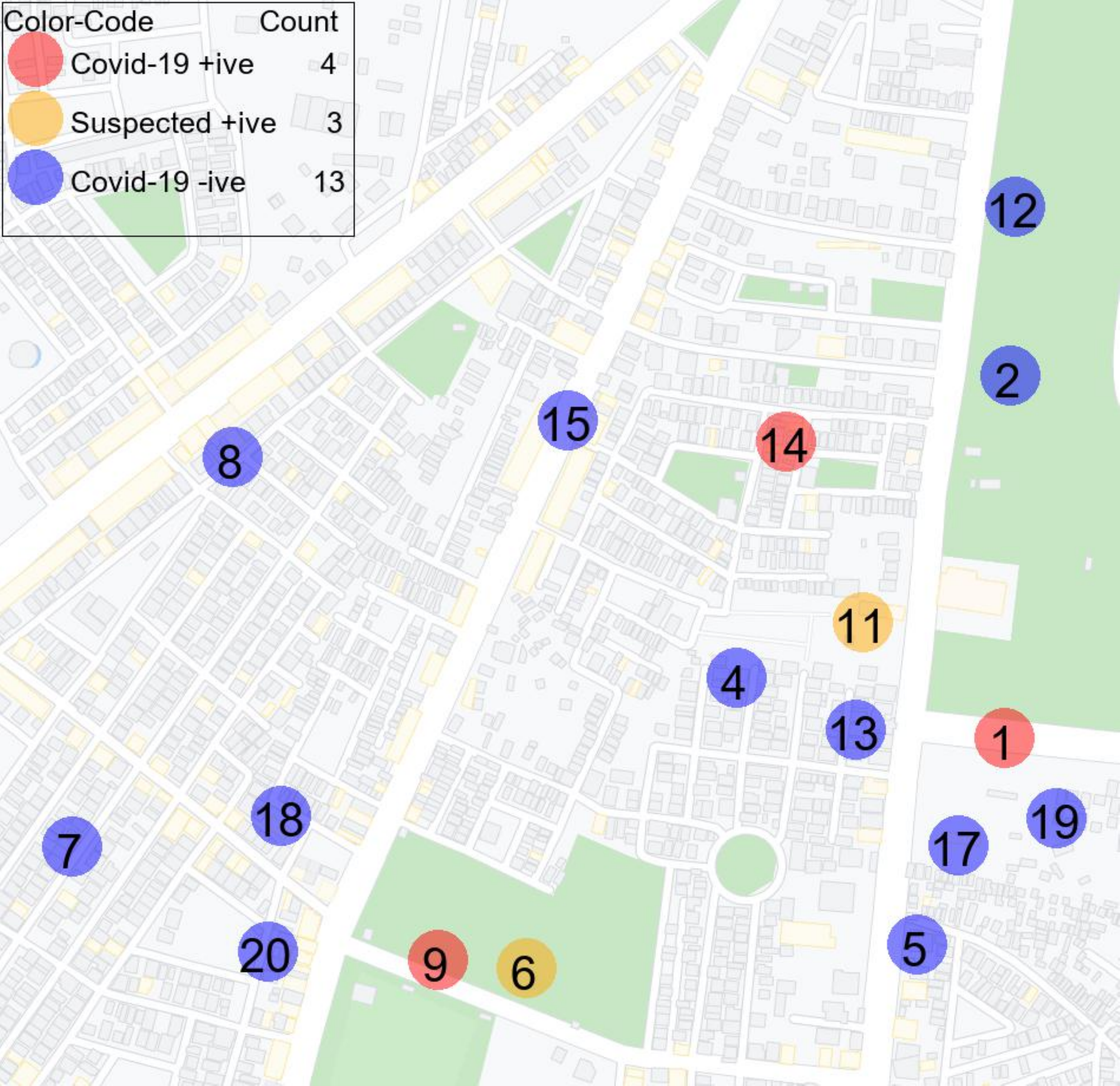}}
    \\
    \subfloat[At time $t_{4}$]{\includegraphics[height=4cm]{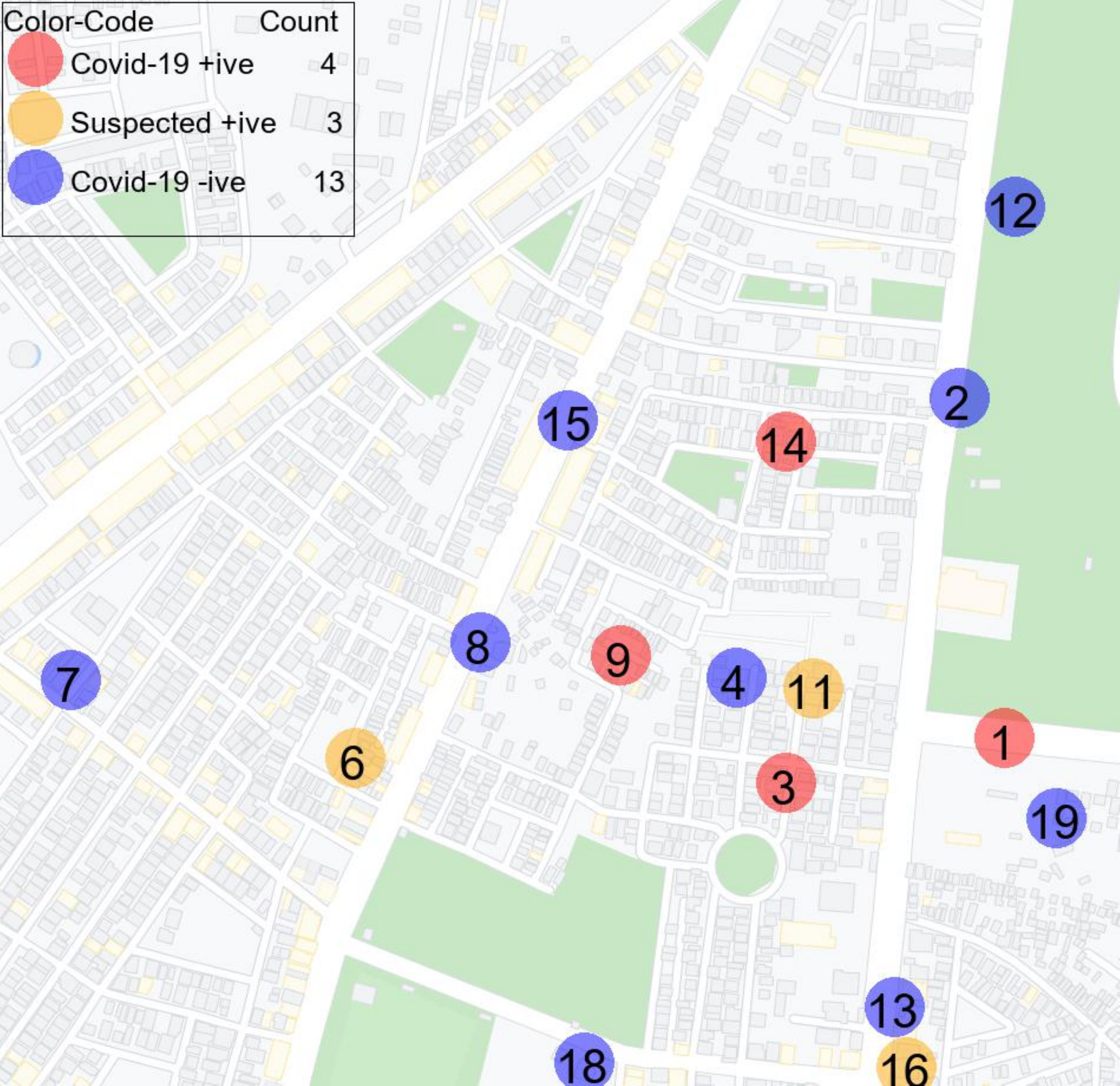}}
    \subfloat[][The runtime of the hotspot detection \\algorithm  wrt different number of users]
{  \includegraphics[width=5cm, height=3cm]{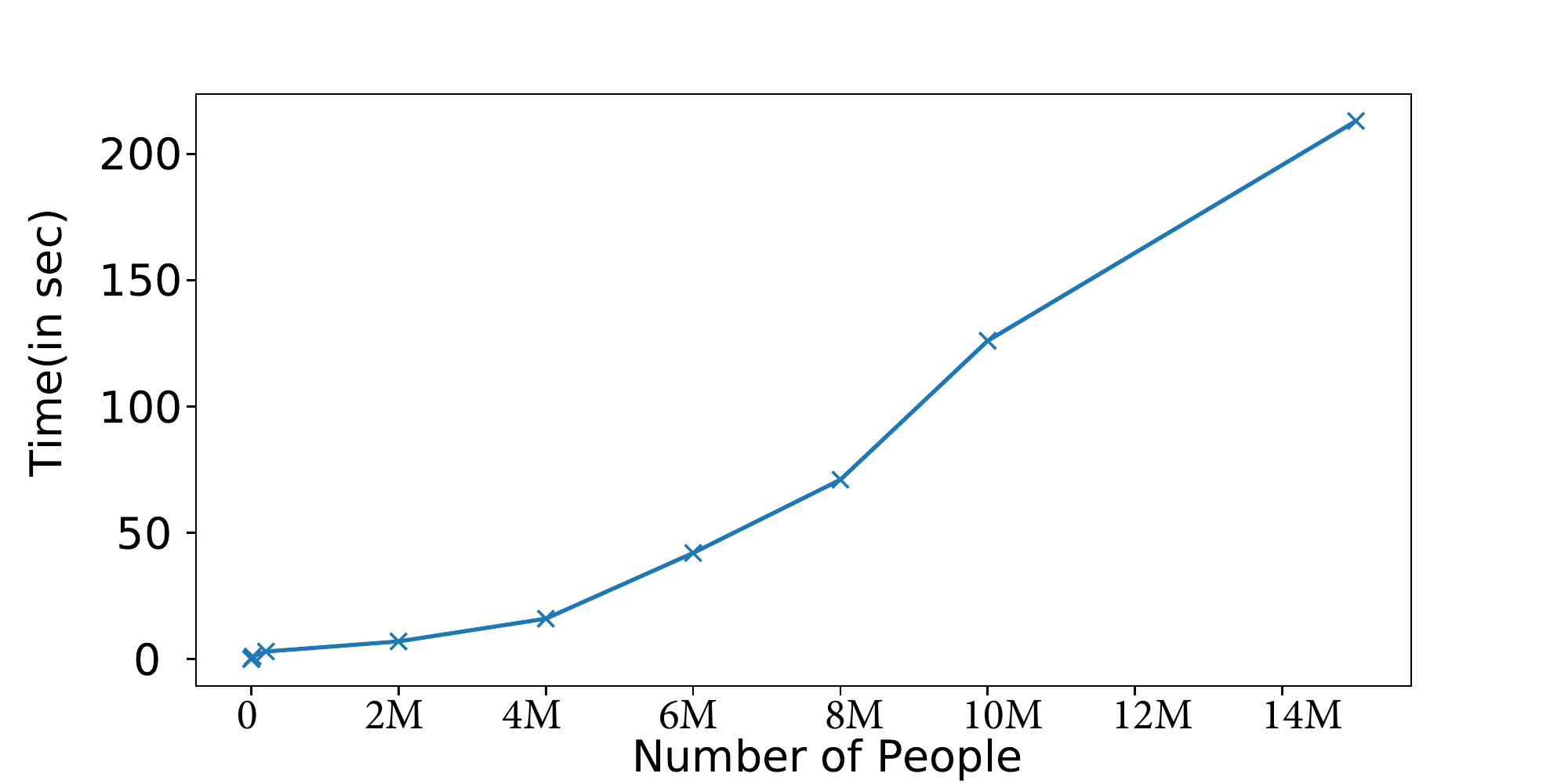}}

    %\hfill
  \caption{Hotspot information at 5 consecutive time periods. Best viewed with zooming the image and in color.}
  \label{hot} 
\end{figure}

To verify the approach discussed in section \ref{hotspotdetection} for hotspot detection, we first test it on 20 users and mark them on google map as shown in fig. \ref{hot}.  All the users, the COVID +ve patients, and the suspected users (users who came in contact with COVID +ve users but still having no symptoms) are plotted on the city map.  The figure also shows that at time $t_0$, users ids 1, 3, 9, and 14 are COVID +ve. The fig. \ref{hot}(b) shows that user 6 who came in contact with user 9 (a COVID +ve), has been suspected and marked with the yellow color.

% \begin{figure}
% \centering
%   \subfloat[ At time $t_{0}$]{\includegraphics[width=0.5\linewidth]{images/hotspot/mapop0.pdf}}
%   \subfloat[ At time $t_{1}$]{\includegraphics[width=0.5\linewidth]{images/hotspot/mapop1.pdf}}
%   \subfloat[ At time $t_{2}$]{ \includegraphics[width=0.5\linewidth]{images/hotspot/mapop2.pdf}}
%   \subfloat[ At time $t_{3}$]{ \includegraphics[width=0.5\linewidth]{images/hotspot/mapop3.pdf} }
%   \subfloat[ At time $t_{4}$]{ \includegraphics[width=0.5\linewidth]{images/hotspot/mapop4.pdf}}
 
% \caption{Hotspot information at 5 consecutive time periods. Best viewed with zooming the image and in color.}
% \label{hot}
% \end{figure}

% \begin{figure*}
% \centering
% \begin{tabular}{cccc}
% \includegraphics[width=0.23\textwidth]{images/hotspot/mapop0.pdf} &
% \includegraphics[width=0.23\textwidth]{images/hotspot/mapop1.pdf} &
% \includegraphics[width=0.23\textwidth]{images/hotspot/mapop2.pdf} \\
% \textbf{(a) At time $t_{0}$}  & \textbf{(b) At time $t_{1}$} & \textbf{(c) At time $t_{2}$}  \\[6pt]
% \end{tabular}
% \begin{tabular}{cccc}
% \includegraphics[width=0.23\textwidth]{images/hotspot/mapop3.pdf} &
% \includegraphics[width=0.23\textwidth]{images/hotspot/mapop4.pdf} \\
% \textbf{(d) At time $t_{3}$}  & \textbf{(e) At time $t_{4}$}  \\[6pt]
% \end{tabular}
% \caption{Hotspot information at 5 consecutive time periods. Best viewed with zooming the image and in color.}
% \label{hot}
% \end{figure*}

Similarly, user 16 becomes susceptible at time $t_2$ and so on. Indeed, the figure shows how the infected/suspected users \emph{may} be moving in the city to provide real-time information of danger zones, benefiting normal users to plan accordingly.  If the number of infected/suspected users in the map relative to the reference point exceeds a certain threshold, then the area around the reference point can be marked as a potential hotspot area. Alternatively, any area comprising a suspected COVID-19 positive patient can be declared as a \emph{potential hotspot}.

One concern related to privacy is that we are marking users at the individual level. To avoid such a thing, we can draw a circle of larger radius such as 10 or 20 meters to conceal the COVID +ve users' identity. Alternatively, we can group nearby users and form large clusters to make the identity indistinguishable at the individual level.  

% \begin{figure}
% \small
% \centering
% \includegraphics[width=6cm, height=3cm]{images/timehot.pdf}
% \caption{The runtime of the hotspot detection algorithm  wrt different number of users}
% \label{timehotspot}
% \end{figure}

To show the scalability of the proposed approach, the plot of runtime with respect to the number of users considered in a hotspot region is shown in fig. \ref{hot}(f). The graph clearly shows that the runtime is following a \emph{near}-linear trend. Therefore, our approach to detect hotspots in real-time may be useful.

% \begin{figure}
% \centering
% \begin{tabular}{cccc}
% \includegraphics[width=0.32\linewidth]{images/hotspot/mapop0.pdf} &
% \includegraphics[width=0.32\linewidth]{images/hotspot/mapop1.pdf} &
% \includegraphics[width=0.32\linewidth]{images/hotspot/mapop2.pdf} \\
% (a) At time $t_{0}$  & (b) At time $t_{1}$ & (c) At time $t_{2}$ \\[6pt]
% \end{tabular}
% \begin{tabular}{cccc}
% \includegraphics[width=0.3\linewidth]{images/hotspot/mapop3.pdf} &
% \includegraphics[width=0.3\linewidth]{images/hotspot/mapop4.pdf} \\
% (d) At time $t_{3}$  & (e) At time $t_{4}$ \\[6pt]
% \end{tabular}
% \caption{Hotspot information at 5 consecutive time periods}
% \label{hot}
% \end{figure}
\begin{figure}
\subfloat[][Route recommended by\\ Google map]{
\includegraphics[width=4cm, height=4.5cm]{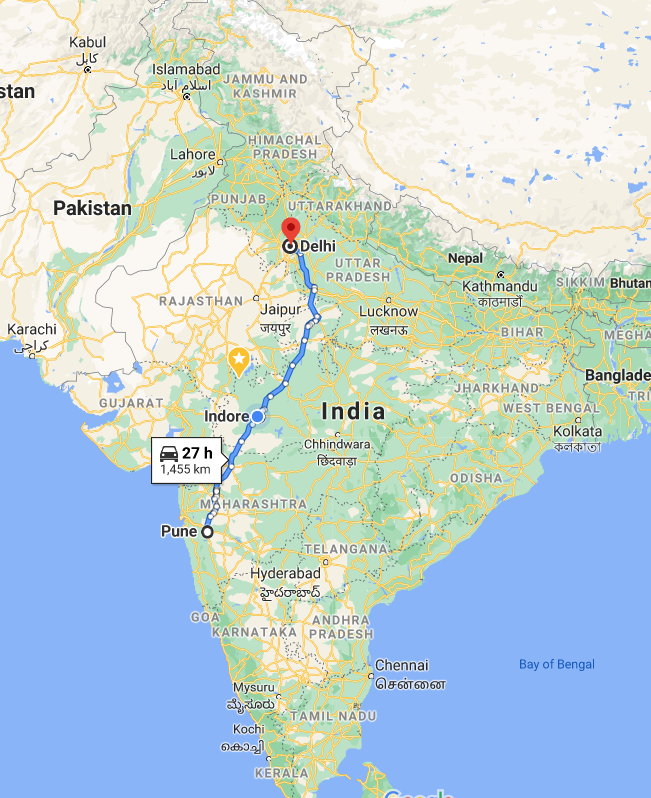}} 
\subfloat[Route recommended by our approach]{
    \includegraphics[width=4cm, height=4.5cm]{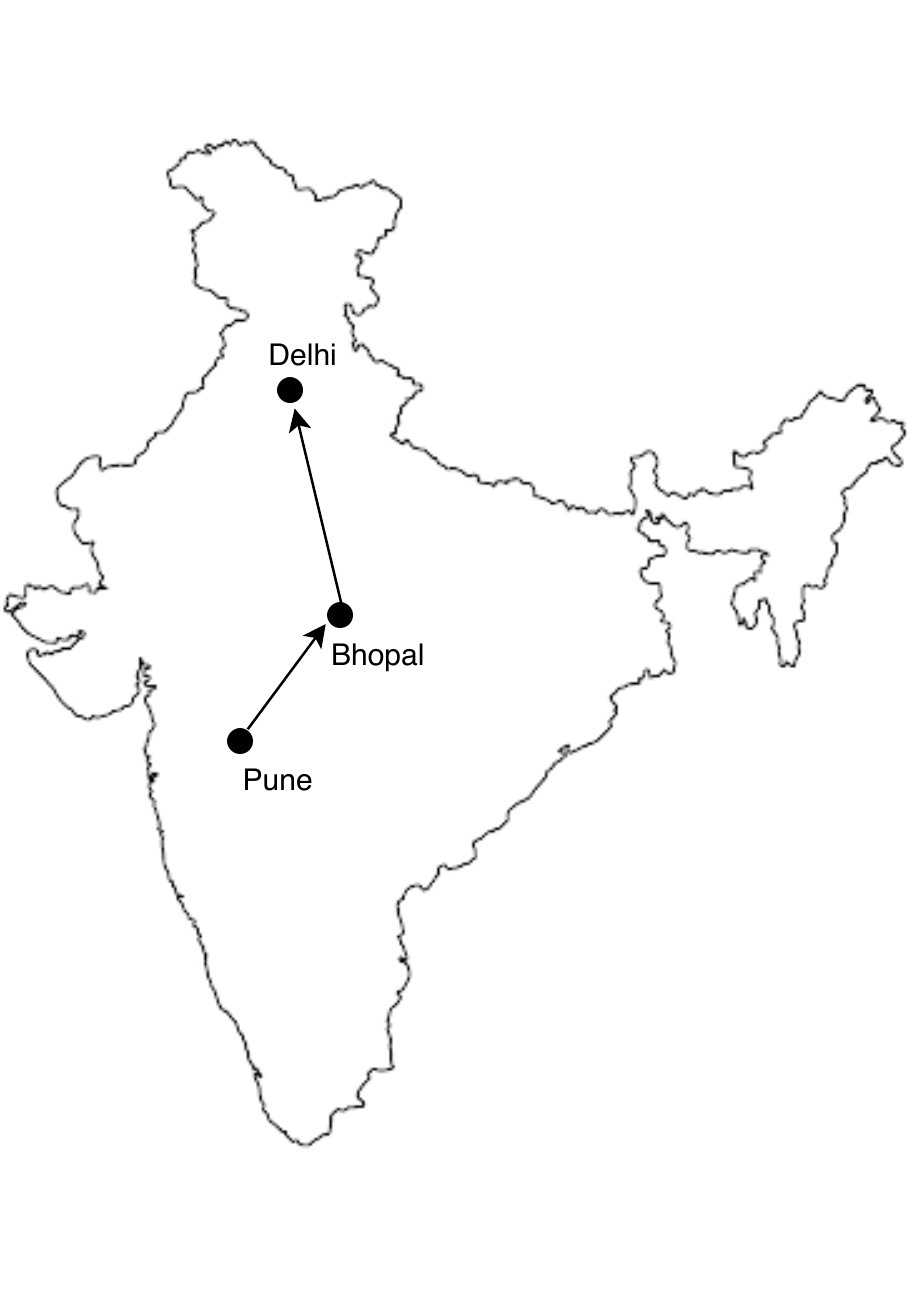}}
\caption{(a) Route recommended by Google map (b) Route recommended by our approach.
We can see that Google map recommends a route from Pune to New Delhi, which passes through Indore city, a hotspot area and not allowed for travel. On the other hand, our approach recommends a route that avoids the hotspot city.}
\label{map}
\end{figure}

% \begin{figure}
% \centering
% \begin{subfigure}
%   \includegraphics[width=0.50\linewidth]{images/s.png}
%   \caption{Route recommended by Google map}
%   \label{fig:Ng1} 
% \end{subfigure}

% \begin{subfigure}
%   \includegraphics[width=0.50\linewidth]{images/map.pdf}
%   \caption{Route recommended by our approach}
%   \label{fig:Ng2}
% \end{subfigure}

% \caption[asds]{(a) Route recommended by Google map (b) Route recommended by our approach.
% We can see that Google map recommends a route from Pune to New Delhi, which passes through Indore city, a hotspot area and not allowed for travel. On the other hand, our approach recommends a route that avoids the hotspot city.}
% \label{map}
% \end{figure}

% \begin{figure}
%   \centering
%   \begin{tabular}{@{}c@{}}
%     \includegraphics[width=.5\linewidth]{images/s.png} \\[\abovecaptionskip]
%     \small (a) Route recommended by Google map
%   \end{tabular}

%   \begin{tabular}{@{}c@{}}
%     \includegraphics[width=.5\linewidth]{images/map.pdf} \\[\abovecaptionskip]
%     \small (b) Route recommended by our approach
%   \end{tabular}

%   \caption{(a) Route recommended by Google map (b) Route recommended by our approach.
% We can see that Google map recommends a route from Pune to New Delhi, which passes through Indore city, a hotspot area and not allowed for travel. On the other hand, our approach recommends a route that avoids the hotspot city.}\label{map}
% \end{figure}

\subsection{Safe Route Recommendation Experiment}
The evaluation of the safe route recommendation approach is done in the following way: 1) The proposed method is compared with an existing route Recommendation system, that is, the Google Map, in terms of the total cost of the path. The total distance from the source to destination constitutes the total cost and whether it is a Hotspot Zone free path or not. The efficacy of the proposed approach is analyzed on a real World dataset, which is extracted from Google Map. For the experiment, we have considered major cities in India (Indore, Bhopal, Chennai, Mumbai, Delhi, Bangalore, Lucknow, Hyderabad, Pune, Kolkata).

We have compared our proposed approach with the current Google map recommendation, considering Indore as a Hotspot City, Source as Pune, and Destination as New Delhi. Google maps recommend the path which passes through the hotspot city(Indore) as shown in Fig. \ref{map}(a) with a total cost of 1455 Km whereas, the path displayed by the proposed algorithm is an alternate route where the intermediate city is Bhopal(non-hotspot city) instead of Indore with a total cost of 1540 Km which is displayed in Fig. \ref{map}(b). Indeed, the path recommended by our approach is not optimal. Still, since our objective is to avoid a hotspot city, our approach is suitable for choosing a safe route rather than an optimal path.

\section{Conclusion and Future Work}
The proposed approach for contact tracing seems plausible based on the initial experiments on simulated data. Safe route recommendation and potential hotspot information further add new features to our method. We are planning to release the app in the future after verification on the real-user study.
% \begin{acks}
% \end{acks}

\newpage
\bibliographystyle{ACM-Reference-Format}
\bibliography{sample-base}

\end{document}